\documentclass[prb,twocolumn,amsmath,amssymb,superscriptaddress,floatfix]{revtex4}

\usepackage[english]{babel}
\usepackage{amsfonts}
\usepackage{graphicx}
\usepackage{times}
\usepackage{soul}

\usepackage{color}
\definecolor{nred} {RGB}{224,0,0}
\definecolor{nblue} {RGB}{28,130,185}
\definecolor{dgreen} {RGB}{78,138,21}

\newcommand{\mat}[1]{\mathrm{#1}}

\begin{document}
%\title{Optical  switching between   electron-electron attraction and repulsion in  a model with nonlinear electron-phonon coupling}
%\title{Optical manipulation of electron - electron interaction in a nonlinearly coupled electron phonon model}
\title{Optical manipulation of bipolarons  in a system with  nonlinear electron-phonon coupling}
%\title{Photo-induced metastable bipolaron due to nonlinear electron-phonon coupling}

\author{K. \surname{Kova\v c} }
\affiliation{J. Stefan Institute, 1000 Ljubljana, Slovenia}

\author{D. \surname{Gole\v z} }
\affiliation{J. Stefan Institute, 1000 Ljubljana, Slovenia}
\affiliation{Faculty of Mathematics and Physics, University of Ljubljana, 1000
Ljubljana, Slovenia}

\author{M. Mierzejewski}
\affiliation{Department of Theoretical Physics, Faculty of Fundamental Problems of Technology, Wroc\l aw University of Science and Technology, 50-370 Wroc\l aw, Poland}

\author{J. \surname{Bon\v ca}}
\affiliation{J. Stefan Institute, 1000 Ljubljana, Slovenia}
\affiliation{Faculty of Mathematics and Physics, University of Ljubljana, 1000
Ljubljana, Slovenia}

\date{\today}
\begin{abstract}

We investigate full quantum mechanical evolution of two electrons nonlinearly coupled to quantum phonons and  simulate the dynamical response of the system subject to a short spatially uniform optical pulse that couples to dipole-active vibrational modes. Nonlinear electron-phonon coupling can either soften or  stiffen the phonon frequency in the presence of electron density. 
%In the atomic limit, both cases lower the energy of the doubly occupied site compared to two singly-occupied ones [D. M. Kennes {\it et al.}, Nat. Phys. {\bf 13}, 479 (2017)]. 
In the former case, an external optical pulse tuned just below the phonon frequency generates attraction between electrons and leads to a long-lived bound state even after the optical pulse is switched off. It originates from a dynamical modification of the self-trapping potential that  induces a metastable state.
%In the case of phonon stiffening a repulsive electrons interaction is always induced. 
By increasing the pulse frequency, the attractive electron-electron interaction changes to repulsive. Two sequential optical pulses with different frequencies can switch between attractive and repulsive interaction. Finally, we show that the pulse-induced binding of electrons is shown to be  efficient also for weakly dispersive optical phonons,  in the presence anharmonic phonon spectrum and in two dimensions.

\end{abstract}  

%\pacs{ } 
\maketitle

\setcounter{figure}{0}

%\section{Introduction}

Research in the field of driven quantum materials is at the forefront of modern solid-state physics. The development of new laser sources opened a new chapter in the field~\cite{hsieh_2017} where we can selectively excite collective degrees of freedom, like lattice, magnetic and electronic excitations, to generate new emergent states of matter~\cite{Giannetti2016,Salen2019,Torre2021}. Among the most prominent examples are  optical manipulation of magnetic order~\cite{Kirilyuk2010,cavalleri_RP_2011,eckstein2015,cavalleri2017}, light-induced non-equilibrium metal--insulator transitions~\cite{cavalleri2007,okamoto2007,cavalleri_SciR2014,georges2014,mihailovic2014}, and optically enhanced transient states displaying superconducting signatures\cite{cavalleri2011,cavalleri_NM2014,cavalleri_Nat2016,demler2016,Giannetti2016,demler2017,freericks2017,georges2017,tohyama2019,galitski2021,cavalleri_PRX2022}.

% Among various approaches to control physical properties of matter\cite{hsieh_2017}, properly   tuned optical  pulses\cite{Giannetti2016,Salen2019,Torre2021} are often  used to alter lattice, magnetic   and electronic properties of complex materials  in order to generate  new emergent states of matter. Among most prominent examples are  optical manipulation of magnetic order, \cite{Kirilyuk2010,cavalleri_RP_2011,eckstein2015,cavalleri2017} light  induced non-equilibrium metal--insulator transitions\cite{cavalleri2007,okamoto2007,cavalleri_SciR2014,georges2014,mihailovic2014}, and optically enhanced transient states displaying superconducting signatures\cite{cavalleri2011,cavalleri_NM2014,cavalleri_Nat2016,demler2016,Giannetti2016,demler2017,freericks2017,georges2017,tohyama2019,galitski2021,cavalleri_PRX2022}

%\cite{cavalleri_2011} Nonlinear coupling between IR and Ramman active phonon to obtain Ionic Ramman scattering.  

%\cite{cavalleri_SciR2014} they investigate the origin of vibrationally controlled Hubbard interaction via an asymmetric holon-doublon  coupling to a nonlinear lattice displacement. Theory is based on the squeezing of the phonon frequency in the presence of the increased phonon density. Also in \cite{cavalleri_PRX2022}. 

Modifying superconducting transition temperature by external stimulus was first shown using microwave radiation, now known as the Wyatt-Dayem effect~\cite{wyatt1966,dayem1967,eliashberg1970}. More recently, it has been suggested that the selective excitation of a system could dramatically enhance the effect on the electronic system with nonlinear lattice couplings. A classic example includes a nonlinear coupling between optically excited infrared active mode inducing a Raman mode distortion~\cite{cavalleri_2011,disa2021,georges2014,subedi2021} with modified electronic properties. The idea was applied to cuprate superconductors~\cite{mankowsky2014}, metal-insulator transition in manganites~\cite{cavalleri2007,esposito2017} and paraelectric-ferroelectric transition in SrTiO$_3$~\cite{subedi2017,nova2019}. An even more interesting class of scenarios explores the role of quantum mechanical fluctuations on pairing from nonlinear phononics\cite{Kiselov2021}. We expect nonlinear electron-phonon coupling in crystals  where light ions are symmetrically intercalated between heavy ions\cite{berciu2014a} like in organic crystals TMTSF$_2$-PF$_6$\cite{kaveh1982}  or TTF-TCNQ\cite{wegner1977,entin1985}.  Recent experimental realization of strong nonlinear electron-phonon coupling was reported in one dimensional ET-F$_2$TCNQ  and identified by the presence of strong second harmonics\cite{kaiser2014,singla2015}. In equilibrium, the nonlinear electron-phonon (EP) coupling leads to light polarons\cite{berciu2014a}  and even more significantly to strongly bound light bipolarons\cite{berciu2014}. Out of equilibrium, the squeezed electronic states due to nonlinear electron-lattice coupling can induce attraction between charge carriers inducing either superconducting~\cite{millis_2017} or insulator-metal transition~\cite{eckstein_2021,sentef2017}. The second class of ideas is based on the parametric resonance effect, where the interplay of driving and lattice non-linearities leads to enhanced electron-phonon interaction and pairing~\cite{demler2017}. However, other studies point out the competition between pairing,  heating or phonon-induced disorder  and, depending on the approximation employed, one or another could prevail~\cite{werner2017,sous_nc_2021}. Therefore, obtaining exact results for a driven system to understand the competition and gauge which approximations are appropriate for these highly excited correlated states would be highly valuable.

In  this Letter, we provide an exact time evolution of a two-electron system coupled non-linearly to lattice distortions and driven by  an external laser pulse, which homogeneously excites dipolar active lattice modes. We show that electronic binding can be dramatically enhanced or reduced depending on the pulse protocol. The electron binding~(or repulsion) remains enhanced due to modified bipolaronic self-trapping leading to a long-lived~(metastable) state  even  after the pulse has been switched off. We show  that the binding takes place only for negative values of non-linear electron-lattice couplings~(frequency softening), which is in contrast with previously applied approximation predicting a sign  independent binding~\cite{millis_2017,demler2017,sentef2017,sous_nc_2021}. Finally, we show that the metastable state slowly   decays  if Einstein phonons acquire a finite bandwidth, still, the binding  remains elevated in comparison to its value before the application of the pulse.

The model under the consideration is given by 
\begin{eqnarray} \label{hol}
\vspace*{-0.0cm}
H_0 &=& -t_\mathrm{el}\sum_{\langle i,j\rangle,s}(c^\dagger_{i,s} c_{j,s} +\mathrm{H.c.}) + {g_2} \sum_{j} \hat n_{j} (a_{j}^\dagger + a_{j})^2+ \nonumber \\
 &+& \omega_0\sum_{j} a_{j}^\dagger  a_{j} + U\sum_{j}n_{i,\uparrow}n_{i,\downarrow}, 
\label{ham}
\end{eqnarray}
where $\langle i,j\rangle$ represents summation over nearest neighbors,   $c^\dagger_{j,s}$ and $a^\dagger_{j}$ are electron and phonon creation operators at site $j$ and spin $s$, respectively,  $\hat n _{j} = \sum_s c^\dagger_{j,s}c_{j,s}$ represents the  electron density operator, $t_\mat{el}$ the nearest-neighbor electron hopping amplitude, and $\omega_0$ denotes the Einstein phonon frequency. From here and on we set $t_\mat{el}=1$.  The second term in  Eq.~(\ref{ham})  represents a quadratic EP coupling, see the Supplemental information (SM)\cite{supp} for physical realizations, and the last term   the on--site Coulomb repulsion.

Since  the studied Hamiltonian exists in infinitely dimensional  Hilbert space one needs to single out a subspace that is relevant for the studied problem. Here, we focus on the dynamics of electrons thus the relevant subspace
 contains states where multiple phononic excitations may exist in the close proximity of electrons. More distant (in real space) phonon excitations are discarded since they do not influence the distribution of electrons.
  In order to construct such subspace
we have used a  numerical  method described in detail in Refs.\cite{bonca1,bonca2000,bonca2002,bonca2012,bonca2016,bonca2021} as well as in the SM.\cite{supp} The method contains a single parameter $N_h$ that determines  the maximal distance between electrons as well as the maximal number of phonon excitations. 

\begin{figure}[!tbh]
\includegraphics[width=1.0\columnwidth]{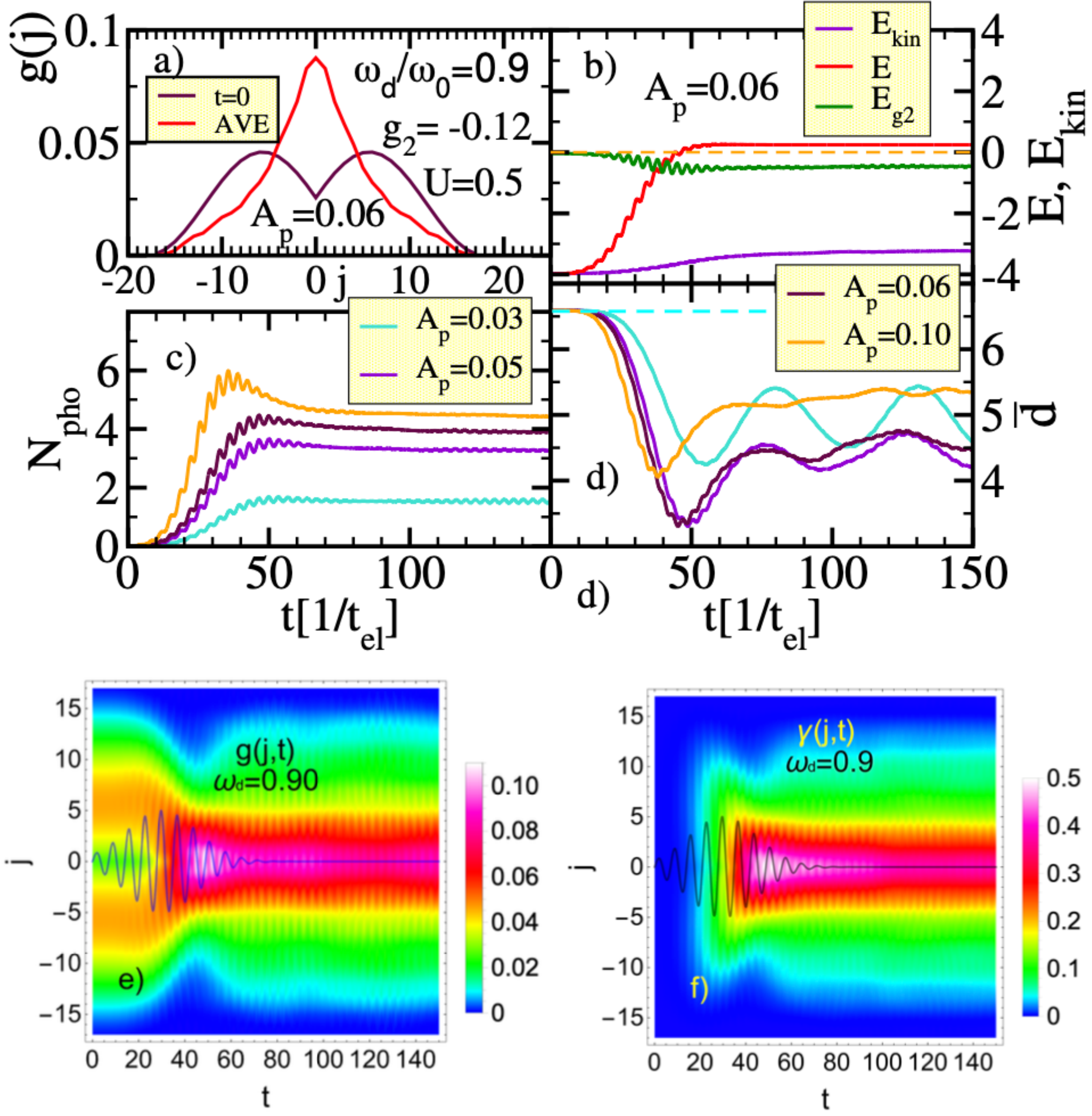}
\caption{ Different expectation values for 1D system:   a) the density--density correlation function in the ground state $g(j,t=0)$  and time--averaged $\bar g(j)$ in the time interval [45-150] after the pulse $V(t)$ with amplitude $A_p=0.06$, $\sigma=15$, $t_0=30$,  and frequency $\omega_d/\omega_0=0.9$;   b) the total energy $E(t)=\langle H\rangle_t$, the kinetic energy $E_\mat{kin}(t)=\langle H_\mat{kin}\rangle_t$ and EP coupling energy $E_{g_2}(t)=\langle H_{g_2} \rangle_t$ where  $H_\mat{kin}$ and $H_{g_2}$ represent the first and the second term in Eq.~(\ref{ham}), respectively;  c) the total number of phonon excitations $N_\mat{pho}=\langle\sum_ja^\dagger_j a_j\rangle_t$ at different pulse amplitudes $A_p$;  d) the average particle distance $\bar d$;  e) the density--density correlation function $g(j,t)$;   f) the number of phonons as a function of the inter--electron distance $\gamma(j,t)$; in e) and f) we used $A_p = 0.06$ and the shape of the pulse $V (t)$ is depicted with a blue line, while its
vertical scale is in arbitrary units. In all figures, the driving frequency is $\omega_d/\omega_0=0.9$ and the electron-phonon coupling is $g_2=-0.12$.
%
% In this and subsequent figures unless otherwise specified  we have used the following  parameters of the model: $t_\mat{el}=\omega_0=1$, $U=0.5$,   $g_2=-0.12$ and $A_p=0.06$.    The  Hilbert space was generated with $N_h=17$ that leads $N_\mat{stat}\sim 1.1\times 10^6$ basis states. 
}
\label{fig1}
\end{figure}

In the first part we present results for the one--dimensional (1D) case.
We excite the system by driving an infrared active mode $H(t)=H_0 + V(t) \sum_{j}  (a_{j}^\dagger + a_{j}) $ with a classical uniform AC field  $V(t)=A_p \sin(\omega_d t) \exp\left [ -(t-t_0)^2/2 \sigma^2 \right ]$ that couples uniformly to all lattice displacements and time-propagate the full problem using the standard Lanczos procedure\cite{park1986}.  Our first observable is the time evolution of the density--density operator 
\begin{equation}
\hat g(j)=\sum_i  \hat n_{i,\uparrow} \hat n_{i+j,\downarrow};~~~g(j,t)=\langle \hat g(j) \rangle_t, 
\label{gj}
\end{equation}
where  we use the following notation $\langle \hat A\rangle_t=\langle\psi(t)\vert \hat A \vert\psi(t)\rangle$.  In the case of two electrons with the opposite spins, the operator $\hat g(j)$ is a projector which singles out states for which the distance between both electrons equals $j$ and   the  average distance between electrons can be obtained as $\bar{d}(t) = \sum_j \vert j \vert g(j,t)$. 
% In particular, one finds identities   $\hat g(j) \hat g(l)=\delta_{jl} \hat g(j) $, $\sum_j  \hat g(j)=1$ and $0\le g(j,t) \le 1$.

In Fig.~\ref{fig1} a) we present $g(j,t=0)$ in the initial ground state of the system using a small Coulomb repulsion $U=0.5$ that overcomes a weak phonon-mediated attraction.  Functional dependence of $g(j,t=0)$ is consistent with two electrons at an average distance $\bar d(t=0)\sim 6.5$ as also seen from  Fig.~\ref{fig1} d). Switching on  the pulse $V(t)$ with a characteristic frequency $\omega_d/\omega_0=0.9$ and different amplitudes $A_p$ causes the increase of the total and  kinetic energies,  $E$ and $E_\mat{kin}$, respectively,    and a slight decrease of  EP coupling  energy $E_{g_2}$,  see Fig. ~\ref{fig1} b). For definitions see  the caption of Fig.~\ref{fig1}. The increase of  $E$ is predominantly due to the  increase of the total number of phonon quanta $N_\mat{pho}$, shown in Fig.~\ref{fig1} c). The most notable  effect of the pulse is  a substantial decrease of the average distance $\bar d$ between electrons, seen in Fig.~\ref{fig1} d).
%, in agreement with DecII. 
While the increase  of $N_\mat{pho}$ with increasing $A_p$ up to $A_p=0.08$ is monotonous, the decrease of $\bar d$ is not and the largest drop is achieved around $A_p\sim 0.06\pm 0.02$. The non-monotonous behavior  originates from the competition between the heating effects and the pairing~\cite{werner2017}.
% not captured in any of the above approximations and the first marker of competition between pairing and heating.  
In Fig.~\ref{fig1} e) the time evolution of the $g(j,t)$  under the influence of the optical pulse is presented as a density plot conjointly with the time evolution of the pulse, $V(t)$.  We observe a  distinct increase of the double occupancy, given by  $g(0,t)$ that  peaks around $t\sim 40 $.  The  increased double occupancy persists even long after the pulse has been switched off creating a very long-lived~(metastable) state, which for the case of Einstein phonons persist almost undistorted up to the largest times used in our calculation.  This is consistent also with a time-averaged $\bar g(j)$ as seen in Fig.~\ref{fig1} a) displaying a peak at $j=0$ in a sharp contrast with its value in the ground state $g(j,0)$.  These observations are in a stark contrast to the Floquet-type scenario analyzed in Ref.~\onlinecite{demler2017}, where the attractive interaction is induced only during the pulse and in the following we will show that it originates from photo-modified self-trapping.

%, see SM\cite{supp}. 

%Once again we should comment that this is not expected in the parametric resonance scenario~(DecII) which would lead to attraction only during the pulse.

Now, we will explore how the long-lived state emerges due to the substantial absorption of the total energy predominantly stored in the increased number of phonon excitations.
% It is worth highlighting that an attractive interaction is achieved between two electrons with opposite spins   via a short optical pulse  accompanied by a  substantial absorption of the total energy that is predominantly stored in the increased number of phonon excitations. 
It is thus worthwhile determining the distribution of the number of phonons as a function of the relative distance between the electrons $j$. It is  measured via the time evolution of $\gamma(j,t) = \langle \hat \gamma(j)\rangle_t$ where
\begin{equation}
\hat \gamma(j) =\hat g(j) \sum_la_l^\dagger a_l, \label{gamma}
\end{equation}
describes the total number of phonons in states where  the distance between electrons equals $j$.
As it  is shown in Fig.~\ref{fig1}  f), most  of the excess phonon excitations are absorbed by doubly occupied  states, i.e. $j=0$,  and those where electrons are in close proximity. It seems as if the excess phonon excitations represent the {\it glue} that at least for the given driving frequency $\omega_d/\omega_0=0.9$ provides a self-trapped attractive potential. 
% Note also that $\gamma(j,t)$ obeys the sum--rule: $\sum_j \gamma(j,t)=N_\mat{pho}(t)$. 

Searching further for the origin of the optically induced electron-electron   potential we realize that it can only originate from
the non-linear electron-phonon interaction term in  Eq.~(\ref{ham}), i.e, from the term $H_{g2}={g_2} \sum_{j} \hat n_{j} (a_{j}^\dagger + a_{j})^2$.
 We define an effective  potential  by projecting $H_{g2}$ on a subspace with a specified distance between electrons:
\begin{equation}
\hat v(j)= \hat g(j) H_{g2},
\end{equation}
that yields the time evolution of the effective potential $v(j,t)=\langle \hat v(j)\rangle_t$. This definition is further justified by  the  sum--rule that  gives   the total interacting energy  $\sum_j v(j,t)=E_{g_2}(t)$ shown in Fig.~\ref{fig1} b). 
%%%%%%%%%%% 

\begin{figure}[!tbh]
\includegraphics[width=1.0\columnwidth]{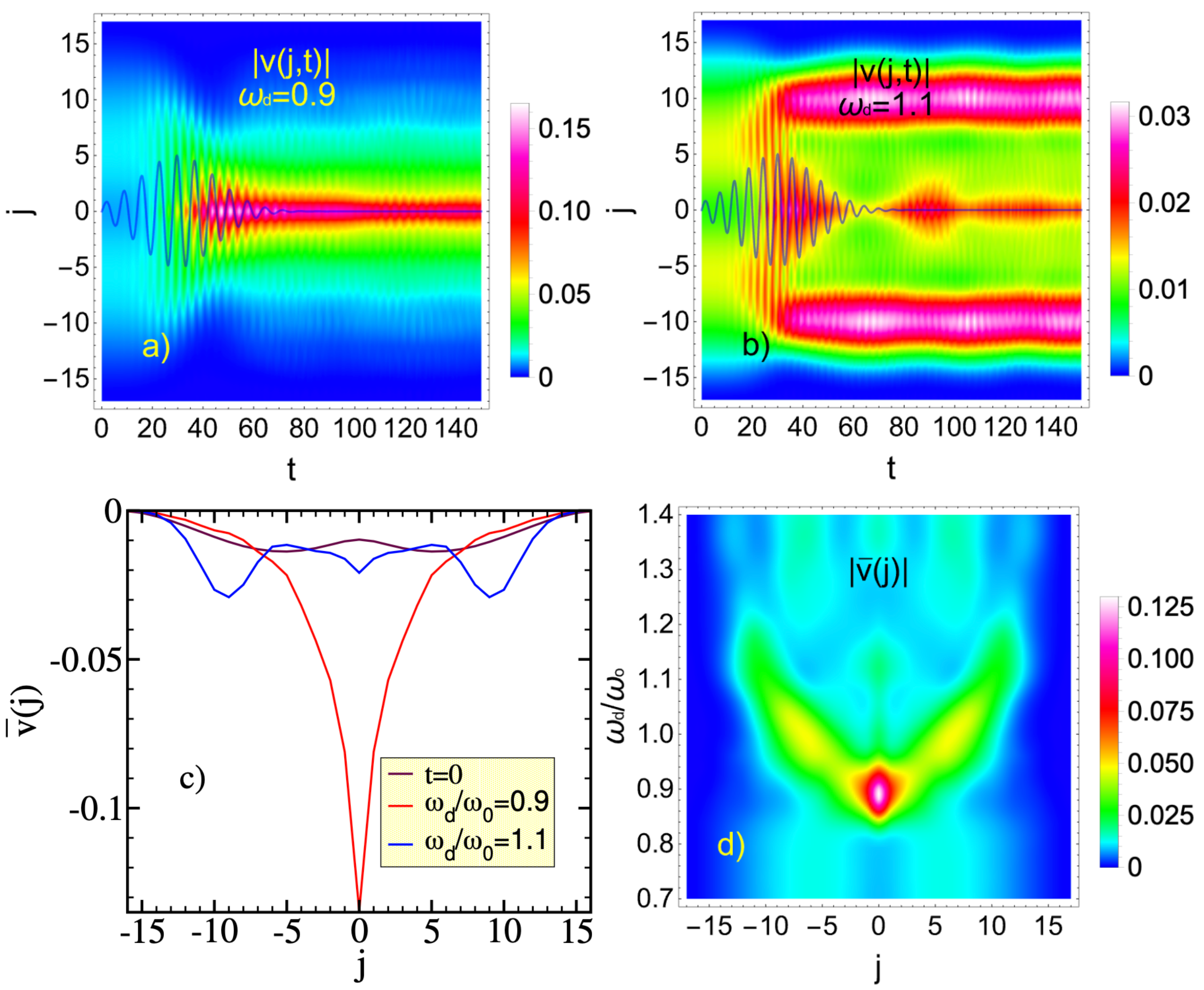}
\caption{  a) and  b) magnitude of the effective  potential $|v(j,t)|$ computed with two distinct driving frequencies $\omega_d/\omega_0=0.9$ and 1.1, respectively. Note that the sign of $v(j,t)$ is strictly negative since $g_2=-0.12$;   c)  the effective potential in the ground state,  $v(j,0)$, and time-averaged $\bar v(j)$ for two distinct $\omega_d$. Time averages were performed in the same interval as in Fig.~\ref{fig1}  a);    d) time-averaged $\bar v(j)$  for different $\omega_d$. We have used the pulse amplitude $A_p=0.06$ while the other   parameters are identical to those used in Fig.~\ref{fig1}. 
}
\label{fig2}
\end{figure}

Motivated by previous Floquet analysis~\cite{demler2017} and a driven atomic limit analysis, see SM\cite{supp}, predicting attractive~(repulsive) electronic interaction for driving below~(above) lattice frequency, we present $v(j,t)$ for two distinct driving frequencies leading to attractive $\omega_d/\omega_0=0.9$ and repulsive $\omega_d/\omega_0=1.1$ interaction for  $g_2=-0.12$, see Figs.~\ref{fig2}   a) and   b). When $\omega_d/\omega_0=0.9$ the pulse generates a pronounced  peak located  at $j=0$, signaling the attractive effective potential that is most negative  for the doubly occupied site. This is also presented in Fig.~\ref{fig2}  c) where we show time averaged $\bar v(j)$.  In contrast, at $\omega_d/\omega_0=1.1$, two minima   appear around $j\pm 10$ that are further apart than the shallow minima in the ground state  at $t=0$, most clearly observed in Fig.~\ref{fig2}  c). This   is consistent with  repulsive interaction considering that our computations are performed on a finite--size system. In Fig.~\ref{fig2}  d) we further investigate  the dependence of the effective potential on $\omega_d$. We present time-averaged  $\bar v(j)$  computed using different driving frequencies $\omega_d$. While the strongest attractive interaction is observed  around $\omega_d/\omega_0=0.9$, with increasing $\omega_d$ the minimum at $j=0$  splits into two separate minima, consistent  with  the onset of a repulsive interaction. Around $\omega_d/\omega_0=1.1$ the separation  of the local  minima reaches  its largest value $|j|\sim 10$.  With  further increase of $\omega_d$ the depth of the local minima diminishes and merges with the background.  The presence of the minima after the pulse is in a clear distinction with the Floquet analysis, where the response is present only during the pulse. The  dynamics of $v(j)$ shows that this interaction originates from the dynamical modification of the trapping potential leading to a long-lived state.

\begin{figure}[!tbh]
\includegraphics[width=1.0\columnwidth]{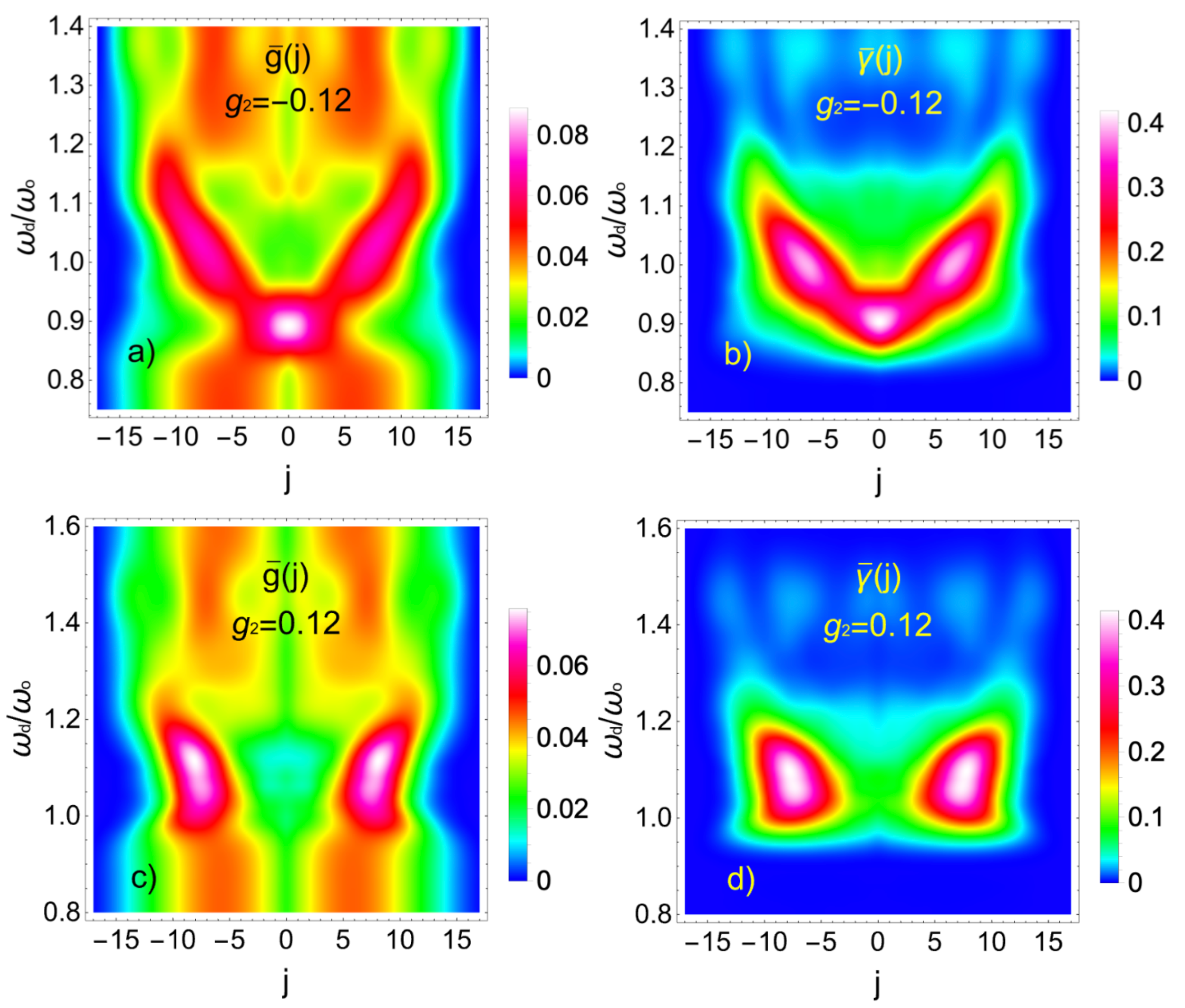}
\caption{  a) and  c)  the time--averaged density--density correlation function $\bar g(j)$  computed using different driving frequencies  $\omega_d$ for $g_2=-0.12$ and 0.12, respectively;  b) and  d) time--averaged phonon distribution  function $\bar \gamma(j)$.  In all cases  the time averages were performed in the same interval as in Figs.~\ref{fig1} a) and ~\ref{fig2} d). We have used the pulse amplitude $A_p=0.06$ while the rest of parameters are identical to those used in Fig.~\ref{fig1}. 
}
\label{fig3}
\end{figure}

% It seems as if only a modest tuning of the driving frequency can change the nature of the optically induced   interaction between electrons.
To obtain a deeper insight  into this phenomenon  we  compute the time--averaged $\bar g(j)$ and the time averaged $\bar \gamma(j)$ and  perform a  scan over  $\omega_d$ at $g_2=-0.12$, see Figs~\ref{fig3}  a) and  b), respectively. For $\omega_d/\omega_0\lesssim 0.8$ and $\omega_d/\omega_0\gtrsim 1.3$,  $\bar g(j)$ resembles its value in the ground state before the pulse has been switched on. This  is consistent with the lack of the absorbed energy from the pulse as seen in  Fig.~\ref{fig3} b) where we observe only a tiny amount of excess phonon excitations in the same  $\omega_d$ regime. Near $\omega_d/\omega_0\sim 0.9$,  $\bar g(j)$ shows a pronounced  maximum due to an increased weight of doubly occupied states ($j=0$) consistent with an attractive electron--electron interaction. With increasing $\omega_d$ the attractive nature of interactions switches  towards a repulsive one as the maximum of $\bar g(j)$ moves  towards larger values of $j$. At $\omega_d/\omega_0 \sim 1.1$ the maximal value of $\bar g(j)$ appears around $j\sim \pm 10$ that  exceeds its average value  in the ground state signaling a strengthening of the  repulsive interaction. 

The evolution of $\bar g(j)$ caused by   changing  driving frequencies $\omega_d$ is closely followed by the evolution of the absorbed    phonon excitation  distributions, $\bar \gamma(j)$, shown in Fig.~\ref{fig3} b).  It is crucial to stress that time  averages are performed in the regime where the optical pulse $V(t)$ that couples to all oscillators  has been switched off. When the pulse is off-resonance,  i.e. when $\omega_d/\omega_0 \gg 1$ or $\omega_d/\omega_0 \ll 1$, the system does not absorb much energy consequently, $\bar \gamma(j)\sim0$ and $\bar g(j)$ remains close to its ground state value. It is worth pointing out that the maximum in the absorbed energy, presented in Fig.~\ref{fig3} b), appears at $\omega_d/\omega_0\sim 0.96$ which is just above the  value  $\omega_d/\omega_0\sim 0.9$ where  maximal attractive interaction is observed.  All this analysis shows that exactly at the resonance the heating effect dominate, but slightly below~(above) the binding~(repulsion) can win and  remarkably leads to a very long-lived state due to the self-trapping mechanism. Remarkably, we observe that the self-traping potential can be reversed by two successive  optical pulses which can switch between attractive and repulsive electron-electron interactions, see SM\cite{supp}.

 % \denis{The induction of the metastable state is the main result of this paper.}

In Figs.~\ref{fig3} c)  and  d) we show results for positive $g_2=0.12$. Absorption of energy from the pulse  appears at higher $\omega_d$ then in the case when $g_2<0$, which is consistent with the increase of the phonon frequency in the presence of finite electron density for  $g_2>0$. In contrast to $g_2<0$ case, the absorption of energy always leads to   repulsive interaction. The observation contrasts with previous analysis~\cite{millis_2017,demler2017,sentef2017,sous_nc_2021} based on the atomic limit or perturbative arguments, which predicted sign-independent pairing. Suppose this observation survives beyond the dilute limit. In that case, it poses a strong constraint on materials where we can expect light-induced long-lived attractive interaction due to the nonlinear electron-lattice couplings. 
% This is further consistent with the average distance $\bar d$, shown in SM\cite{supp},  for   $g_2>0$. 

\begin{figure}[!tbh]
\includegraphics[width=1.0\columnwidth]{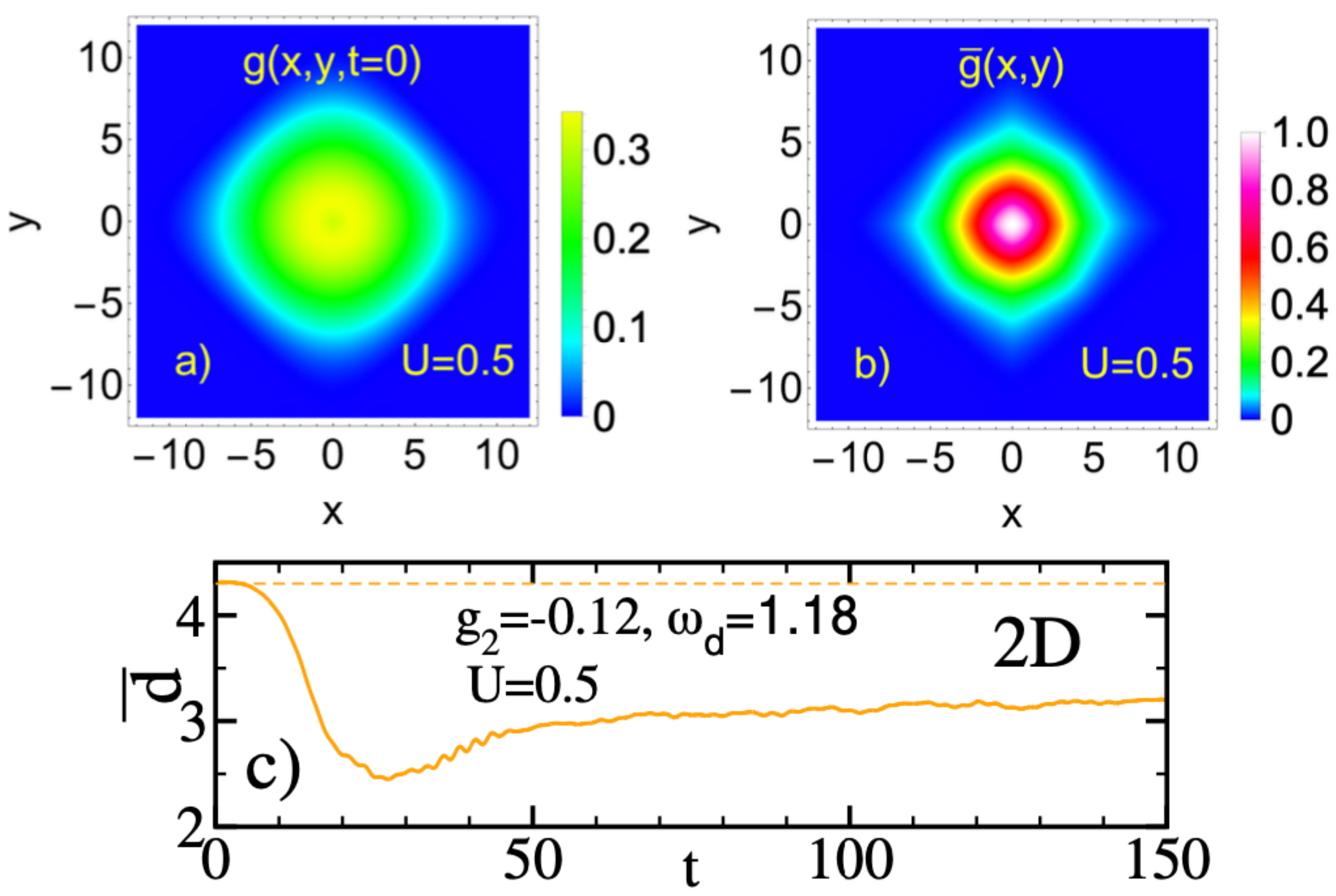}
\caption{ 
 a)   $g(x,y,t=0)$ computed in the ground state at $\omega_0=1, g_2=-0.12$ for the 2D version of the  model in Eq.~\ref{ham};  b)  time--averaged $\bar g(x,y)$ where time averages have been performed in the time interval $t\in[45,150]$; to facilitate comparison between results computed before and after driving, we have rescaled the density plots so that  the largest value of $\bar g(x,y)$ is set to unity; c) the average distance between particles $\bar{d}(t) = \sum_{x,y} \sqrt{x^2+y^2} g(x,y,t)$. We have used $N_h=12$ while  the size of the Hilbert  space was $N_\mat{st}^\mat{2D}\sim 2.6\times 10^6$ and the rest of parameters of the model are the same as in the 1D case. For the driving field we have used the following parameters $\omega_d/\omega_0=1.18$,  $A=0.1$ , $\sigma=15$, $t_0=30$.
}
\label{Fig4}
\end{figure}

Now, we address how robust is the metastable state to various perturbations. In SM\cite{supp}, we show the resilience to the presence of the linear EP coupling term, phonon dispersion, and anharmonic effects on the phonon spectrum.  An important remaining question is whether  the optically induced attraction survives in higher dimensions and we will demonstrate it for the  two dimensional (2D) case of Eq.~\ref{ham}. 
% Last, we present results for the  two dimensional (2D) case of Eq.~\ref{ham} to demonstrate that the optically induced attraction between electrons survives in higher dimensions.  
The model can describe nonlinear coupling to phonon modes that are perpendicular to the plane. 
% Our calculation is again based on the same basic principle, as in Ref.~\cite{bonca1,bonca2000}. 
The system is defined on an infinite 2D plane, taking into account translational symmetry while the maximal distance between electrons is given by $N_h$ and the maximal number of phonons in the system is $N_h-1$~\cite{bonca1,bonca2000}.

Fig.~\ref{Fig4}a) shows the ground state density--density correlation function \mbox{$g(x,y,t=0)$, where} $g(x,y,t)=\sum_{x',y'}\langle\hat n_{(x',y'),\uparrow}\hat n_{(x'+x,y'+y),\downarrow} \rangle_t$ at $g_2=-0.12$.  Due to the presence of nonzero $U=0.5$ there is a shallow local minimum at  the center, presenting the doubly occupied site;  $x=y=0$.  Note also that $g(x,y,t)$ is normalized, {\it i.e.} $\sum_{x,y}g(x,y,t)=1$, it therefore represents the probability for a state where the relative position of  the electrons is given by $(x,y)$.  In Fig.~\ref{Fig4}b)  we present time--averaged $\bar g(x,y)$ after driving. The size of the bipolaron  as a result of driving shrinks while the probability for double occupation  increases. This is more quantitatively shown in Fig.~\ref{Fig4}c) where the average distance $\bar d(t)$ shows a substantial decrease during as well as after the driving.

% In contrast, when $\omega_d\sim\omega_0$ the system can absorb energy from the puls in a range of $\omega$

% In the Supplemental material we present $g(j,t)$ and $\gamma(j,t)$ obtained with the $\omega_d/\omega_0 \sim 1.1$\cite{supp}.

% The increase in $E$ is predominantly due to the increase of phonon energy $E_\mat{pho}=\omega_0 N_\mat{pho}$. 

In conclusion, we have performed numerically exact time evolution of a bipolaron problem coupled by a non-linear electron-phonon interaction. When the electron-lattice interaction leads to phonon softening $g_2<0$, a properly tuned uniform optical pulse that couples to dipole-active lattice vibrations may optically induce either attractive or repulsive interaction between electrons.
Here the primary mechanism originates from strong dependence of the effective phonon frequency on the local density of electrons\cite{millis_2017} so that an appropriately tuned pulse excites phonons corresponding to specific configurations of electrons. 
In 1D the  strongest  attractive interaction appears when pulses are tuned slightly below the Einstein phonon frequency $\omega_0$. This suggests that a softening of the lattice vibration due to an increased electron density and double occupancy plays an essential role in the appearance of the attractive potential between electrons. In contrast, a driving frequency slightly exceeding $\omega_0$ generates repulsive interaction. In both cases, the effects of optically induced interactions persist long after the pulse has been switched off.   Since the energy of the phonon subsystem depends on the density of electrons, both subsystems build an effective trap potential that mutually  stabilizes the spatial configurations of electrons and phonons. We have demonstrated that optically induced interaction  survives under various perturbations of the original Hamiltonian, such as: the introduction of  weakly dispersive phonons, anharmonic effects on phonon frequency spectrum, and  the introduction of the liner EP coupling. The mechanism is stable also  in two and possibly also in higher dimensions.  An important future problem is extending the bipolaron problem to finite doping to understand if we can induce coherence between these highly excited composite particles and to explore  the experimental consequences of such states.

\begin{acknowledgments}
J.B., D.G. and K.K. acknowledge the support by the program No. P1-0044 and No. J1-2455 of the Slovenian Research Agency (ARRS).  J.B.  acknowledge discussions with S.A. Trugman, A. Saxena and support from  the Center for Integrated Nanotechnologies, a U.S. Department of Energy, Office of Basic Energy Sciences user facility and Physics of Condensed Matter and Complex Systems Group (T-4) at Los Alamos National Laboratory. 
M.M. acknowledges support by the National Science Centre, Poland via Project No. 2020/37/B/ST3/00020.
\end{acknowledgments}

\bibliography{manuaomm}
\end{document}

%\end{document}

\title{Optical manipulation of bipolarons  in a system with  nonlinear electron-phonon coupling}

\author{K. \surname{Kova\v c} }
\affiliation{J. Stefan Institute, 1000 Ljubljana, Slovenia}
\affiliation{Faculty of Mathematics and Physics, University of Ljubljana, 1000
Ljubljana, Slovenia}

\author{D. \surname{Gole\v z} }
\affiliation{J. Stefan Institute, 1000 Ljubljana, Slovenia}
\affiliation{Faculty of Mathematics and Physics, University of Ljubljana, 1000
Ljubljana, Slovenia}

\author{M. Mierzejewski}
\affiliation{Department of Theoretical Physics, Faculty of Fundamental Problems of Technology, Wroc\l aw University of Science and Technology, 50-370 Wroc\l aw, Poland}

\author{J. \surname{Bon\v ca}}
\affiliation{J. Stefan Institute, 1000 Ljubljana, Slovenia}
\affiliation{Faculty of Mathematics and Physics, University of Ljubljana, 1000
Ljubljana, Slovenia}

\maketitle

\section{Numerical method} % (fold)

The method is based on a construction of  basis states for the many-body Hilbert space that can be written as $\vert j_1,j_2; \dots, n_m, n_{m+1},\dots,\rangle$,  where the spin up and down electrons are on sites $j_1$ and $j_2$, and there are $n_m$  phonons on site $m$. A functional  subspace is constructed iteratively beginning with an initial state where both electrons are on the same site with no phonons and applying the sum of operators $H_\mat{el} + H_\mat{EP}$ ($H_\mat{el}=\sum_{j,s}(c^\dagger_{j,s} c_{j+1,s} +\mathrm{H.c.})$ and $H_\mat{EP}=\sum_{j} \hat n_{j} (a_{j}^\dagger + a_{j})$)    $N_h$  times taking into account  the full translational symmetry.
The constructed  Hilbert space defined on the one-dimensional chain allows only a finite  maximal distance of a phonon quanta from the   doubly occupied site, $L_\mathrm{max_1}=(N_\mathrm{h}-1)/2$,  a maximal distance between two electrons $L_\mathrm{max_2}=N_\mathrm{h}$, and a maximal amount of phonon quanta at the doubly occupied site $N_\mathrm{phmax}=N_\mathrm{h}$. The number of generation $N_h$ is therefore our convergence parameter and by converging in it one can consider the solution numerically exact. The number of states in the many-body Hilbert space $N_\mat{st}$  in the 1D case increases with $N_\mat{h}$ exponentially as  $N_\mat{st}^\mat{1D}\sim 6.4\times 2^{1.02 N_h}$ and in the 2D case as: $N_\mat{st}^\mat{2D}\sim 3.6\times 3^{1.02 N_h}$. Note that in the limit of large $N_h$ the two respective Hilbert spaces would grow as $N_\mat{st}^\mat{1D}\sim 2^{N_h}$ and 
$N_\mat{st}^\mat{2D}\sim 3^{N_h}$.

\section{Invariability of optically induced interaction against different perturbations}

\subsection{Linear electron--phonon coupling term}

\begin{figure}[!tbh]
\includegraphics[width=1.0\columnwidth]{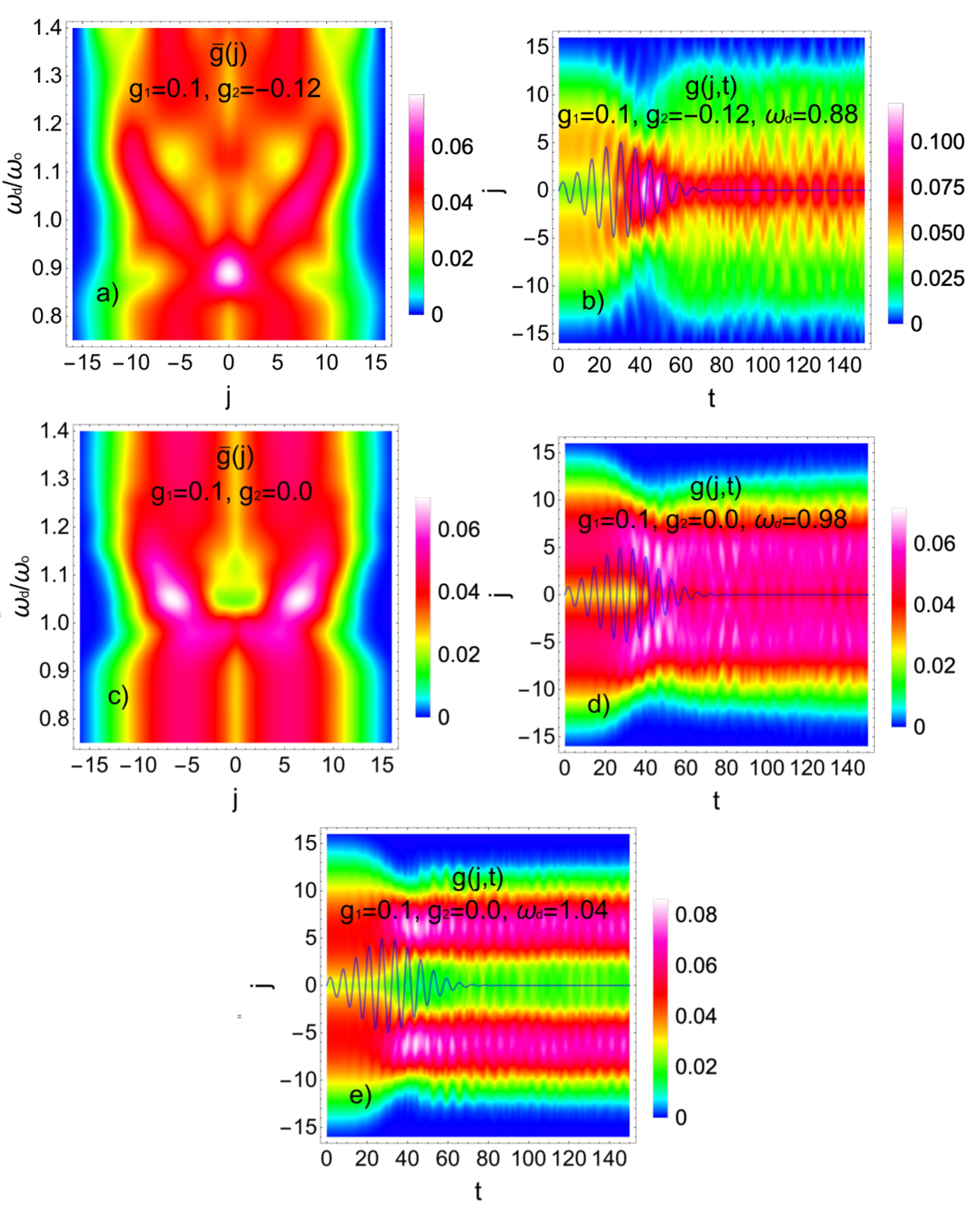}
\caption{  a) and  c)  the time--averaged density--density correlation function $\bar g(j)$  computed using different driving frequencies  $\omega_d$, linear coupling term   $g_1=0.1$,  and two different values of $g_2=-0.12$ in a) and $g_2=0.0$ in  c); in    b),  d), and e)  we present the time evolutions of $g(j,t)$ for parameters, presented in a) and   c) while choosing  driving frequencies $\omega_d/\omega_0=0.88$, 0.98 and 1.04 where maximal attractive or repulsive interaction is expected.  In  a) and c)  the time averages were performed in the same interval  as in Fig.~1  of the main text. We have used the pulse amplitude $A_p=0.06$, $U=0.5$, and $N_h=16$.
%     while the rest of parameters except for  $g_1$  are identical to those used in Fig.~3 a).
}
\label{Fig5S}
\end{figure}

We next demonstrate that the optically induced attraction survives also in the presence of the linear EP coupling, nevertheless, it does not occur without the presence of an anharmonic interaction.
To this end, we test the influence of a linear EP coupling term $H_1 = g_1\sum_j \hat n_j(a_j^\dagger + a_j)$ on the dynamic response of $H_0$ as studied in the main text.  In  Fig.~\ref{Fig5S} we show results obtained after the time propagation with the Hamiltonian $\tilde H(t) = H_0 + H_1 + V(t)\sum_j(a_j^\dagger + a_j)$.   In  Fig.~\ref{Fig5S} a) we first present time--averaged $\bar g(j)$ obtained using different $\omega_d$ at $g_2=-0.12$ and $g_1=0.1$.  We compare results obtained at finite $g_1$  to those when $g_1=0.0$, as shown in  Fig.~3 a) of the main text.  Close to $\omega_d=0.9$ we observe an attractive electron--electron interaction that crosses over to   repulsive  with increasing $\omega_d$, nevertheless, the crossover  is not as  pronounced as in the $g_1=0$ case.  A closer inspection of the time propagation at $\omega_d=0.88$ , as presented in   Fig.~\ref{Fig5S} b), reveals pronounced oscillations of the metastable bound state. This is in contrast with results for $g_1=0.0$,  presented in Fig.~1 e) of the main text, where oscillations are much less pronounced. 

We next set $g_2=0.0$ and by investigating results presented in  Figs.~\ref{Fig5S} c), d), and e)  we consider  whether only the  linear EP coupling term is sufficient to induce dynamically generated interaction between electrons. In comparison between  Fig.~\ref{Fig5S} a), of SM and  Fig.~3 a) of the main text,  we do not observe a clear signal consistent with the optically induced attractive interaction, see also  Fig.~\ref{Fig5S} d). Instead, around $\omega_d/\omega_0=1.04$, presented in  Fig.~\ref{Fig5S} e), a signature of a repulsive interaction is found.

\subsection{Significance of the anharmonic effects in the phonon spectrum.}

\begin{figure}[!tbh]
\includegraphics[width=1.0\columnwidth]{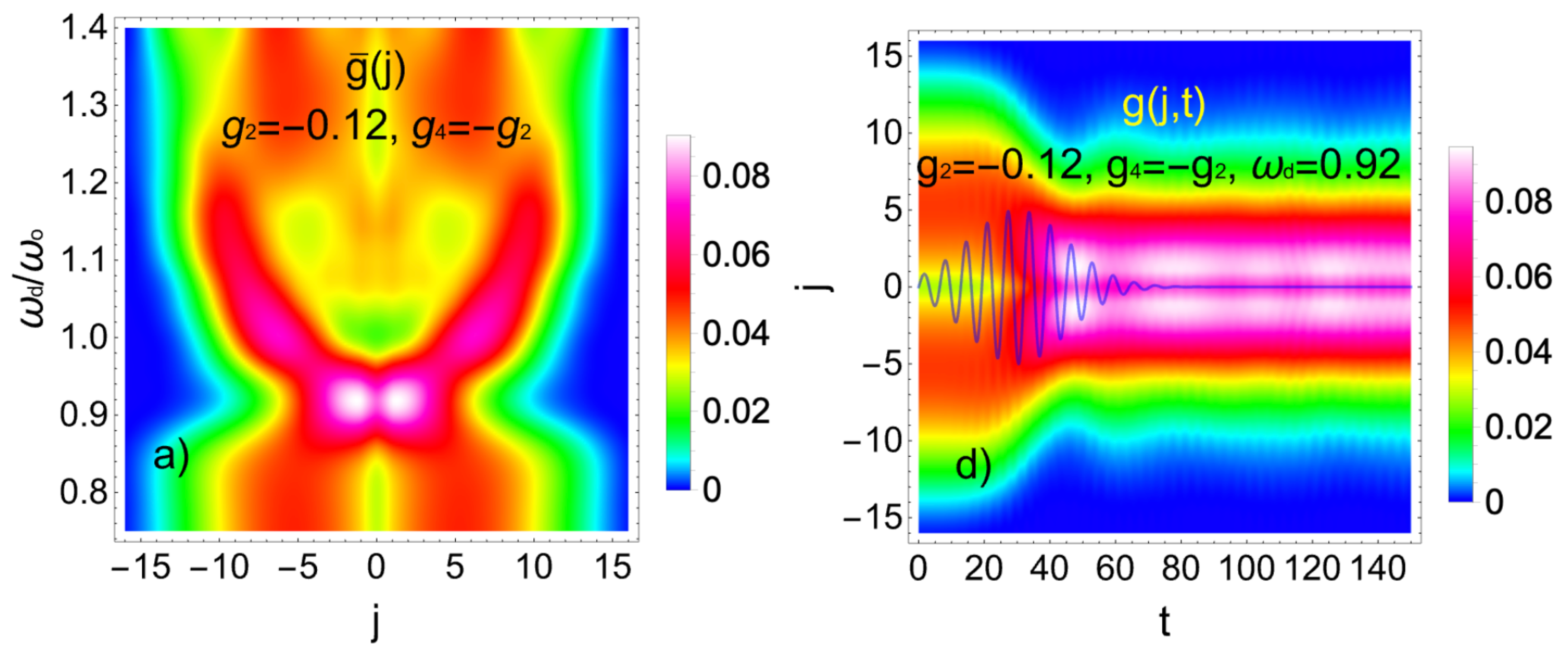}
\caption{  a)  the time--averaged density--density correlation function $\bar g(j)$  computed using different driving frequencies  $\omega_d$, and a renormalized anharmonic phonon spectrum computed  at $g_4=0.12$; in    b) we present the time evolutions of $g(j,t)$ at   driving frequency $\omega_d/\omega_0=0.92$.  In all cases  the time averages were performed in the  same interval as in Fig.~1 of the main text. We have used the pulse amplitude $A_p=0.06$, $U=0.5$, and $N_h=16$. }
%     while the rest of parameters except for  $g_1$  are identical to those used in Fig.~3 a).
\label{Fig6S}
\end{figure}

In the main text we have considered  the  anharmonic EP coupling, however, for the sake of simplicity, the phonon spectra were assumed to be the same as in the case of a simple harmonic oscillator. 
Below, we demonstrate that our conclusions concerning the optically induced attraction hold true also beyond this simplification. 
To this end we first  solve the eigenproblem for the anharmonic oscillator: \mbox{$H_4 = \alpha [ \omega_0 a^\dagger a + g_4 (a^\dagger + a)^4 ]$.}
It order to disentangle the anharmonic effects from a trivial shift of the first excited state, the energy spectrum of $H_4$ is renormalized by a constant $\alpha$. The value of $\alpha$  is set by the condition
 $ \epsilon_1-\epsilon_0=\omega_0$, where  $\epsilon_0$ and $\epsilon_1$  are,   respectively, the ground state and the first excited state energies of $H_4$.  Finally, we perform the time evolution of $H(t) =  H_0 + V(t)\sum_j(a_j^\dagger + a_j)$, where the original phonon spectrum on each site: $0, \omega_0, 2\omega_0, 3\omega_0, \dots$ is replaced by  the renormalized anharmonic spectrum of $H_4$. For example, in the case when $\omega_0=1$ and $g_4=0.12$, the renormalized  spectrum is given by:  $0 , \omega_0, 2.23 \omega_0, 3.61\omega_0,  \dots$.  As observed in Figs.~\ref{Fig6S} a) and b), the dynamically induced electron electron interaction remains robust upon   the anharmonic modification of the phonon spectrum. At resonant driving frequency a  metastable bound bipolaron is formed that  consist of two electrons  located predominantly on adjacent  sites. 

\subsection{Dispersive phonons}
\begin{figure}[!tbh]
\includegraphics[width=1.0\columnwidth]{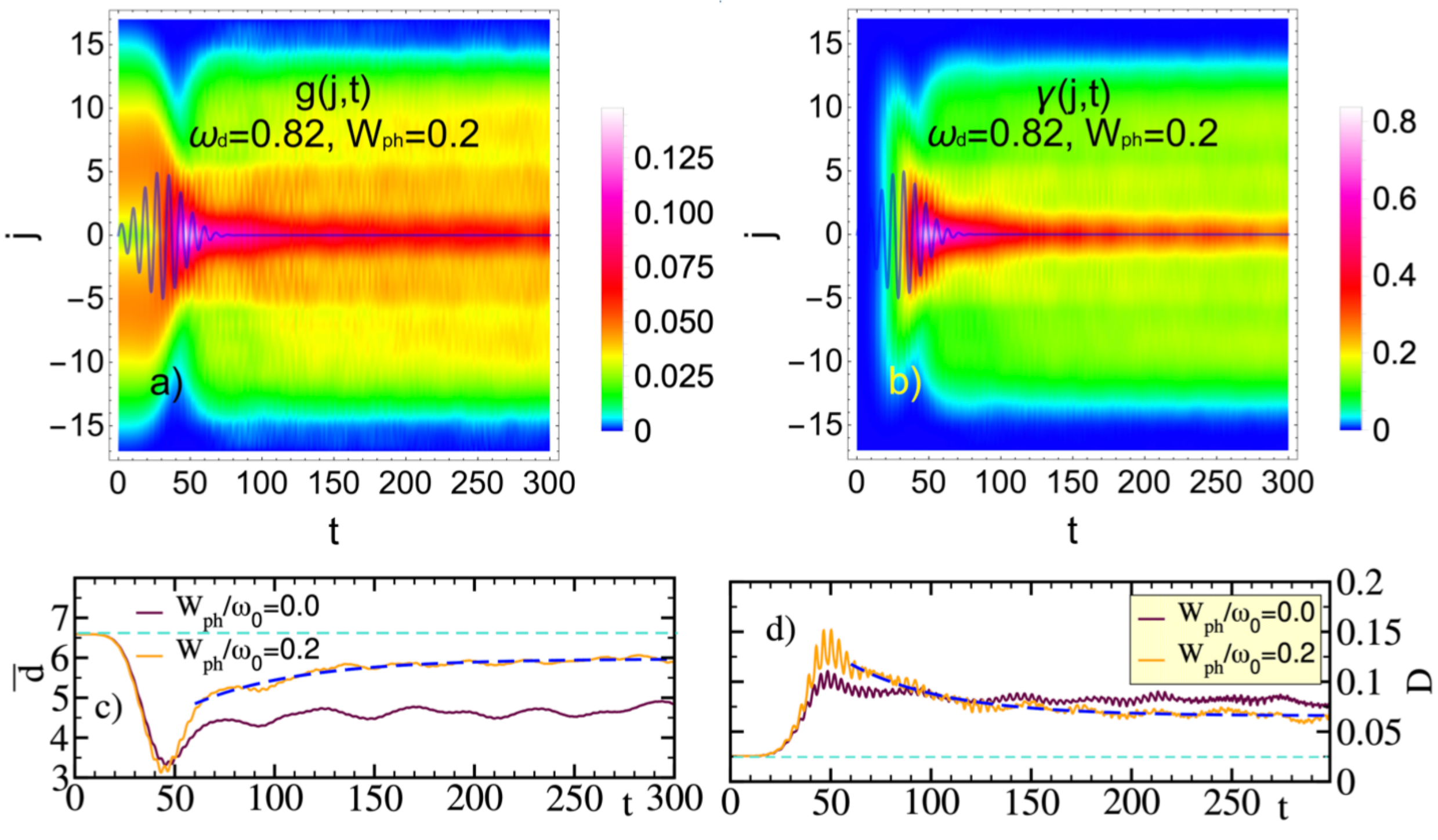}
\caption{ Correlation 
 a) $g(j,t)$  and   b) $\gamma(j,t)$ for dispersive optical phonons using $W_\mat{ph}=0.2$ and optimal driving frequency  $\omega_d/\omega_0=0.82$ that at chosen $W_\mat{ph}$ yields   maximal attractive interaction;  c) and  d) present comparison of $\bar d(t)$ and $N_\mat{pho}(t)$, respectively. Exponential fits (blue dashed lines) in  c) and  d), of the form 
$A(t)={\cal A}\exp(-t/\tau)+A_{\infty}$,
 yield $(\tau_{\bar d},{\bar d_\infty})\sim (53,6.0)$ and $(\tau_{D},{D_\infty})\sim(48,0.07)$, respectively.
The other parameters are identical to those used in Figs.~1  e) and  f) in the main text. 
}
\label{ Fig7S}
\end{figure}

We explore the long-lived stability of the optically driven attraction  between electrons as seen in Fig.~1 e) in the main text, against the introduction of optical phonon dispersion. In Figs.~\ref{ Fig7S}  a) and  b) we show $g(j,t)$ and $\gamma(j,t)$ obtained from a modified hamiltonian  $H = H_0 -W_\mat{ph}/4 \sum_j \left ( a_j^\dagger a_{j+1} + H.c\right )$. The second  term introduces dispersion among optical phonons with the  bandwidth  $W_\mat{ph}$. While the initial decrease of $\bar d$ during the pulse, shown in  Fig.~\ref{ Fig7S} c),    is comparable to $W_\mat{ph}=0$  case, we observe a slow relaxation towards $\bar d_\infty <\bar d_0$ with a relaxation time $\tau_{\bar d}\sim 50\gg 1/W_\mat{ph}$ as the pulse is switched off. In contrast, the double occupancy $D_t=\langle\hat n_\uparrow \hat n_\downarrow\rangle_t$, presented in Fig.~\ref{ Fig7S} d),  shows a larger increase at finite $W_\mat{ph}$ during the pulse, followed by a slow decrease with a relaxation time $\tau_D\sim\tau_{\bar d}$  towards $D_\infty>D_0$ where $D_0$ and $\bar d_0$ represent their respective values in the equilibrium.  Even though the  introduction of finite $W_\mat{ph}$ results in the initial decrease   of $D(t)$ after the pulse, the value of $D_\infty$ is consistent with the attractive interaction surviving the introduction of   $W_\mat{ph}$.

\subsection{Next-nearest-neighbor electron hopping}
\begin{figure}[!tbh]
\includegraphics[width=1.0\columnwidth]{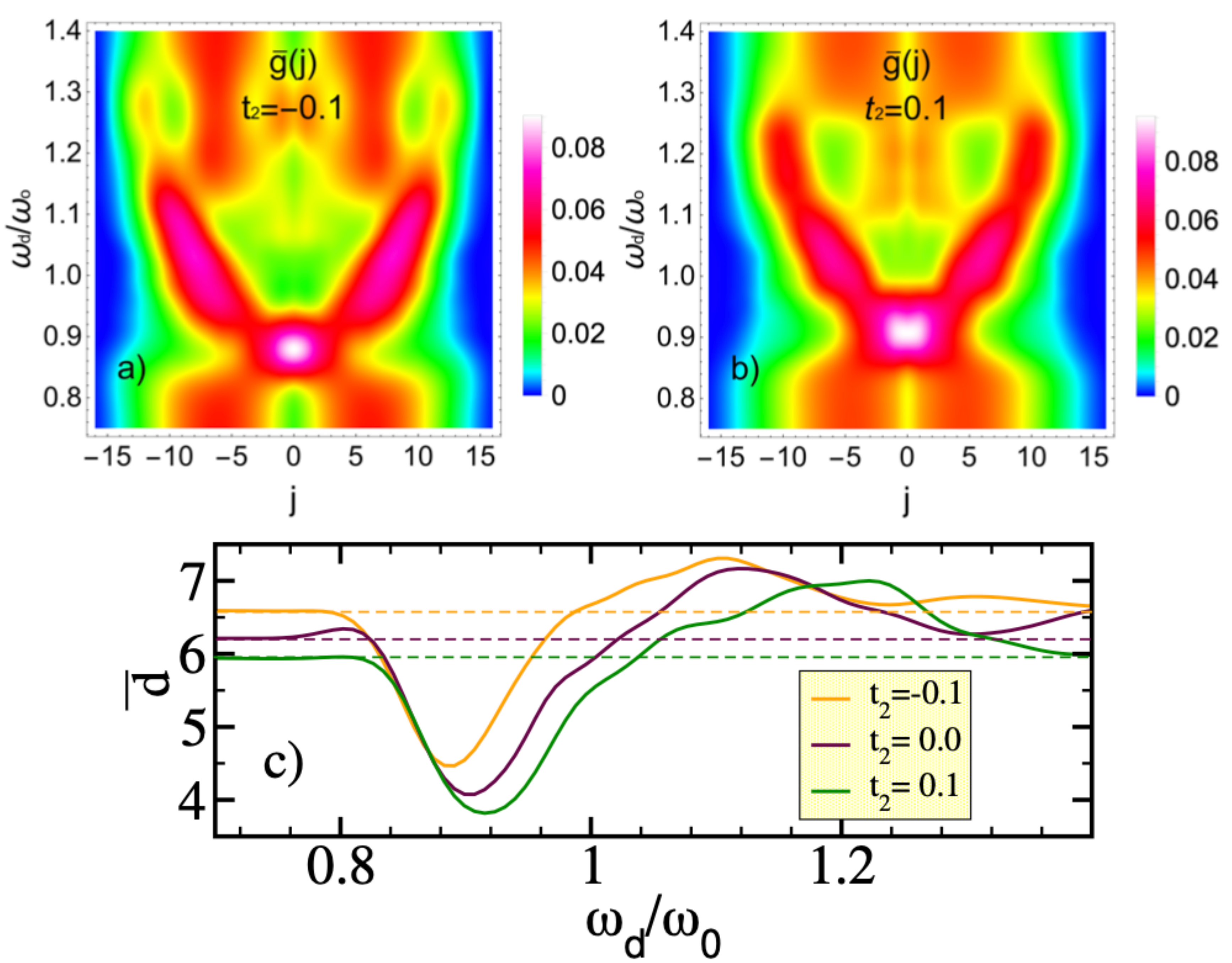}
\caption{  a) and b) $\bar g(j)$ using different values of $t_2=-0.1$ and 0.1, respectively; c) time-averaged average distance $\bar d$ vs. the driving frequency for different values of $t_2$.  
The rest of parameters, $\omega_0=1.0, g_2=-0.12,$ and $U=0.5$,   
are identical to those used in Fig.3a) of the main text. Time averages were performed after the pulse has been switched off. Hilbert space was generated using $N_h=16$. 
}
\label{Fig8S}
\end{figure}

We explore the influence of the next-nearest-neighbour electron hopping term. The modified Hamiltonian is given by: 
$H = H_0 +t_2\sum_{i,s}(c^\dagger_{i,s} c_{i+2,s} +\mathrm{H.c.})$, results are presented in Fig.~\ref{Fig8S}.  The addition of  next-nearest-neighbour hopping brings about  only small changes to the optically generated interaction between electrons. There is a visible shift  in the optimal driving frequency of the optical pulse $\omega_d$ that generates strongest attractive interaction. In addition, a positive value of $t_2=0.1$ slightly increases  attractive interaction in the ground state, while the effect of $t_2=-0.1$ is  opposite, see Fig.~\ref{Fig8S}c). This is best seen comparing different  $\bar  d$ for $\omega_s/\omega_0\lesssim 0.75$ that are nearly identical to those in the ground state since the system does not absorb energy in the low $\omega_d$ regime.  The effect that $t_2$ has on the ground state properties is reflected also  in the optically excited states since the time-averaged values of $\bar g(j)$ show smaller average distance between electrons for $t_2>0$ in comparison to its opposite sign.  Results should also be compared with Fig. 3a) of the main text.

\section{Driving with $\omega_d/\omega_0=1.1$} % (fold)
\begin{figure}[!tbh]
\includegraphics[width=1.0\columnwidth]{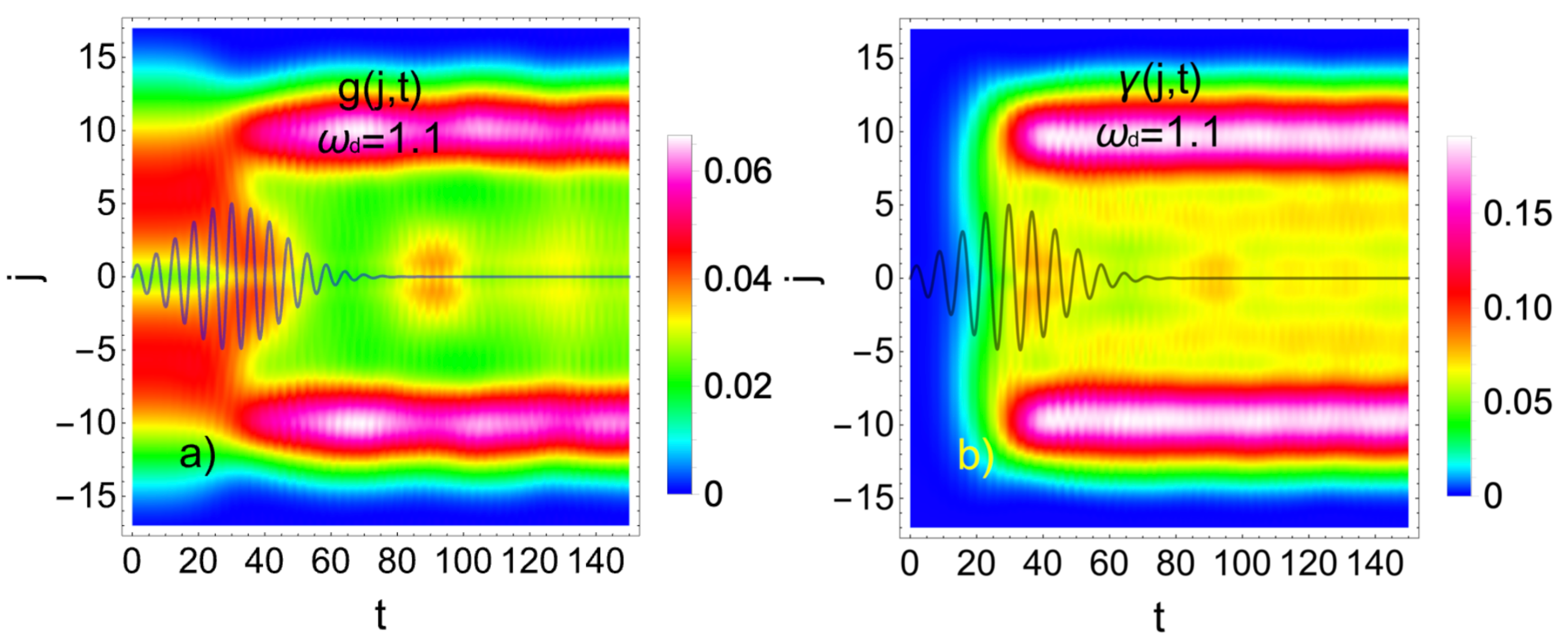}
\caption{  Time--dependent correlation functions at $\omega_d/\omega_0=1.1$. a)  the density--density correlation function $g(j,t)$; b)   the number of phonons as a function of the inter--electron distance   $\gamma(j,t)$.  In both cases the shape of the pulse $A(t)$  with  $\omega_d/\omega_0=1.1$ is shown, its vertical scale is  in  arbitrary units. The rest of parameters is the same as used to generate Fig.~1(e) and (f). 
}
\label{fig1S}
\end{figure}

\label{rep}
In Fig.~\ref{fig1S} we present the time evolution of $g(j,t) $ and $\gamma(j,t)$ under the influence of the optical pulse with the driving frequency $\omega_d/\omega_0=1.1$ that generates repulsive interaction  between the electrons. In comparison to the case presented in Fig.~1e) of the main text where the central  driving frequency is $\omega_d/\omega_0=0.9$, the  maximum of $g(j,t)$ is around $j\sim 0$, in Fig.~\ref{fig1S} a) the peak in $g(j,t)$  is around $j\sim 10$. In addition, the increase  of $\gamma(j,t)$ is more evenly spread to $j\not = 0$ in Fig.~\ref{fig1S} b) than in Fig.~1(f) of the main text . 

\begin{figure}[!tbh]
\includegraphics[width=1.0\columnwidth]{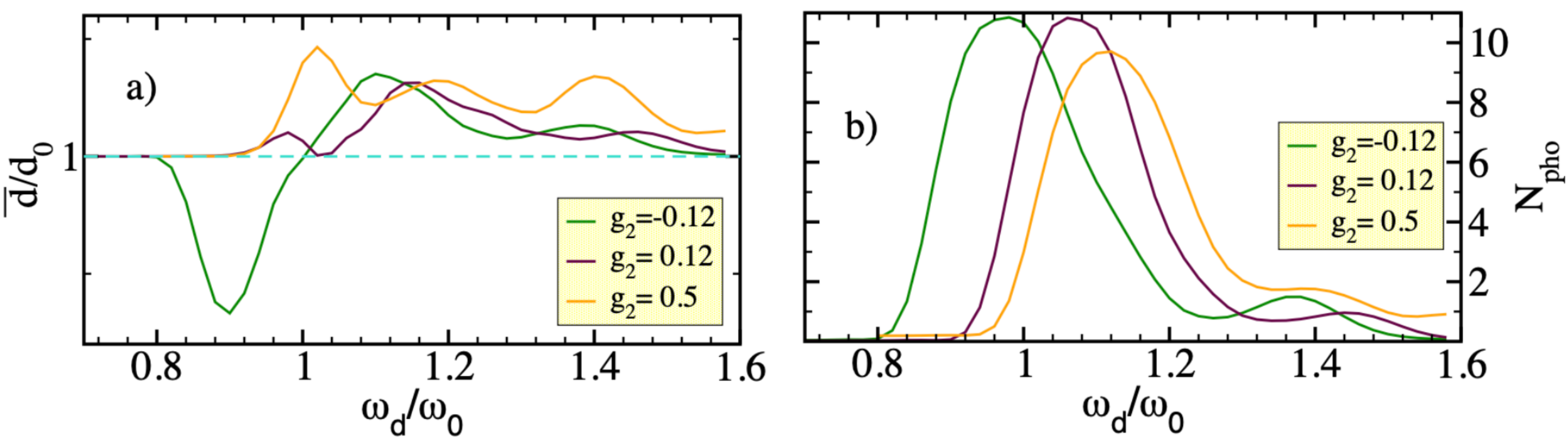}
\caption{ The time--averaged correlation functions computed at $g_2=\pm0.12$, and 0.5   for different driving frequencies $\omega_d$. Time averages are defined as in Fig.~3 of the main text; a) represents the  time--averaged distance relative to its ground state value $d_0$, $\bar d/d_0$ and b) the  total number of phonons $N_\mat{pho}$.  The rest of parameters are identical to those used in Fig.~(3) of the main text. 
}
\label{fig2S}
\end{figure}

\section{Response of the system  for $g_2>0$}
%\label{gplus}
In Fig.~\ref{fig2S} we analyse  the response of the system to different driving frequencies in the case of positive quadratic EP coupling $g_2=0.12$ and 0.5.  This analysis has been stimulated by the results presented in Ref.~\cite{millis_2017} where authors analyse the atomic limit of the model in Eq. ~(1). They compute the effective electron-electron interaction $U^*$  of the model by comparing energies of zero, one and two electrons on the atomic site with an equal number of phonon excitations $N_\mat{pho}$ that gives:
\begin{equation}
U^*=U - (N_\mat{pho}+{1\over 2})\omega_0\left[ 2 \sqrt{1+4g_2/\omega_0} -1 -\sqrt{1+8g_2/\omega_0}\right] .
\label{ren}
\end{equation}
The second term in Eq.~\ref{ren} is negative for any finite value of $g_2 > -\omega_0/8$, nevertheless,  its value is  asymmetric around $g_2=0$. For example, in the case of $g_2=-0.12$, one obtains similar renormalization $U^*$ as for $g_2=0.5$. This suggests an inquiry   whether a positive $g_2$ that gives similar renormalization in the atomic limit also leads to attractive interaction when subject to an optical pulse with a well-tuned frequency. 

%In Fig.~\ref{fig2S} we show results of time-averaged $\bar g(j)$ and $\bar \gamma(j)$ obtained using identical numerical procedure as used for Fig.~(3) of the main text. 

%In contrast  to results for $g_2=-0.12$, presented in Fig.~(3) a) and c) of the main text, results in Figs.~(2) a) and c) show no attractive interaction at any value of $\omega_d$. Instead, for $\omega_d/\omega_0\gtrsim 1.0$ results are consistent with repulsive interaction.  The only similarity between the two cases is  in 
%comparison of Figs.~(3) d) of the main text and Fig.~\ref{fig2S} d) where $N_\mat{pho}$ shows a very similar dependence on $\omega_d$ apart for a shift to larger $\omega_d$. This points to a similar total energy absorption   from the pulse but its distribution relative to the electron position $\bar \gamma(j)$ as displayed in  Figs.~(3) b) of the main text and Fig.~\ref{fig2S} b) shows a stark  contrast. While for $g_2<0$  $\bar \gamma(j)$ peaks at the electron position, at $j=0$, for $g_2>0$  $\bar \gamma(j)$  shows a local minimum  at $j=0$.  

%Next, we investigate whether   an attractive induced interaction between exists for $g_2>0$.  
In Fig.~\ref{fig2S}(a) and (b) we show time-averaged $\bar d$ and $N_\mat{pho}$ as functions of $\omega_d$ for $g_2=-0.12$ and two positive values $g_2=0.12$  and 0.5. For $g_2>0$  $\bar d$ always  increases, consistent with the repulsive interaction, providing that  $\omega_d$ is chosen such that the  system absorbs energy from the pulse that leads to   the increase of $N_\mat{pho}$. When comparing results for $g_2=-0.12$ and +0.12 we observe a very similar increase of $N_\mat{pho}$ with the only observable difference in the shift of the $g_2=0.12$ result towards  slightly larger $\omega_d$.  In contrast, for $g_2=-0.12$,  $\bar d$ is consistent with attractive interaction for $0.8\lesssim \omega_d/\omega_0\lesssim 1.0$ and repulsive for $1.0\lesssim \omega_d/\omega_0\lesssim 1.6$.  It is worth stressing that in the latter case, the deep in $\bar d$ is reached around $\omega_d/\omega_0=0.9$ while the peak in $N_\mat{pho}$ appears around $\omega_d/\omega_0=0.96$. 

\section{Finite Hilbert space analysis}

\begin{figure}[!tbh]
\includegraphics[width=1.0\columnwidth]{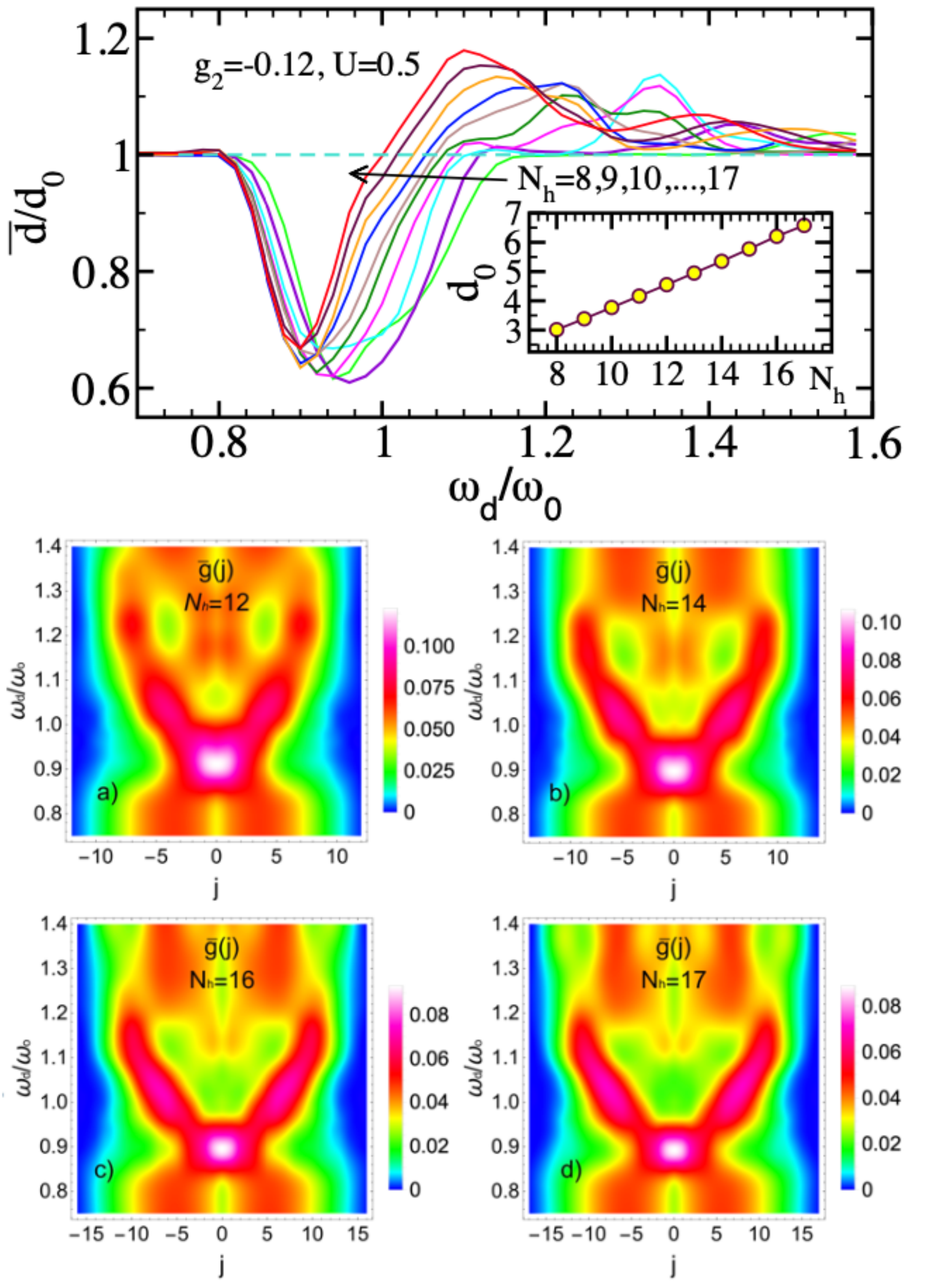}
\caption{ Top figure: the time--averaged $\bar d/d_0$ for different sizes of the Hilbert space ranging from $N_\mat{st}^\mat{1D}\sim 1600$ for $N_h=8$,   up to $N_\mat{st}^\mat{1D}\sim 1.1\times 10^6$ for $N_h=17$. $d_0$ represents  the average distance in the ground state. Note that  $N_h$ represents the maximal distance between electrons and the maximal number of phonon quanta in the limited Hilbert space. The inset shows  the average distance $ d_0$ in the ground state before the pulse has been switched on.  The rest of the parameters are identical to those used in Fig.~(3) of the main text. 
From a) through d) we show the evolution of the time-averaged density-density correlation functions $\bar g(j)$ with increasing number of basis states generated by $N_h=12, 14, 16$ and 17 that lead  to $N_\mat{st}=31064, 130843, 542591$ and $1.1\times 10^6$ basis states, respectively. }
\label{fig3S}
\end{figure}

In Fig.~\ref{fig3S} we show how the time--averaged $\bar d$ changes with increasing  size of the Hilbert space $N_\mat{st}^\mat{1D}$  as a function of $\omega_d$. Note first that $N_\mat{h}$ that sets the number of many-body Hilbert states   $N_\mat{st}^\mat{1D}\sim 6.4\times 2^{1.02 N_h}$   also  defines  the maximal allowed distance between electrons.    The decrease of $\bar d$ around $\omega_d/\omega_0\sim 0.9$ seems to be well converged in terms of the optimal value of $\omega_d$ that yields maximal drop of $\bar d$ as well as its relative change $\bar d/d_0$. In contrast, the increase of $\bar d/d_0$ that indicates repulsion at $\omega_d/\omega_0 > 1$ grows in size.  This is consistent with the repulsive nature of interaction taking into account  limited distance between the electrons due to the restricted Hilbert space. In the inset of the top panel of Fig.~\ref{fig3S} we show $d_0$ in the ground  state for different $N_\mat{h}$. Even though the average distance $d_0$ before the application of the pulse  increases with the system size due to finite $U=0.5$, the relative decrease  $\bar d/d_0$ seems to converge with increasing $N_\mat{h}$  signaling the effectiveness of the optically  induced attractive interaction. 

In Figs.~\ref{fig3S}a) through d) we show the evolution of time-averaged $\bar g(j)$ with the system size. It seems that by increasing the system size the switching between the attractive and repulsive interaction becomes even more pronounced. 

\section{Double pulse}

\begin{figure}[!tbh]
\includegraphics[width=1.0\columnwidth]{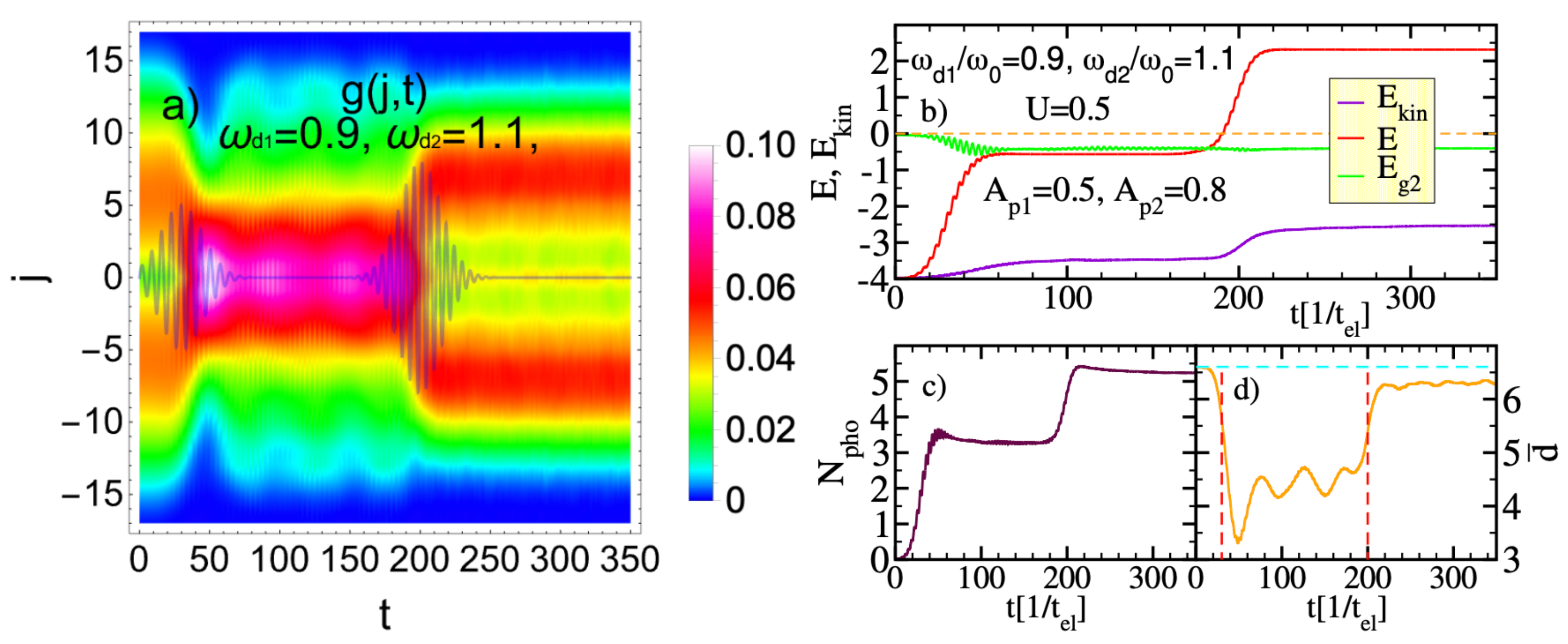}
\caption{ 
 a) $g(j,t)$ in the case of two successive optical pulses with two different $\omega_d$ and $A_p$ as denoted in the legend;  b) different energies as defined in the caption of Fig.~(1) of the main text;  c) and  d),  $N_\mat{pho}$ and $\bar d$, respectively. In  d) vertical lines present times $t_1=45$ and $t_2=200$ of the optical pulses given by   $V(t)=A_{p1} \sin(\omega_{d1} t) \exp\left [ -(t-t_1)^2/2 \sigma^2 \right ] + A_{p2} \sin(\omega_{d2} t) \exp\left [ -(t-t_2)^2/2 \sigma^2 \right ]$ as also  shown with a grey line in  a). The other  parameters are identical to those used in Fig.~(1) of the main text. 
}
\label{Fig4S}
\end{figure}

In Fig.~\ref{Fig4S} we present a simulation of a system with $g_2=-0.12$ subject to two consecutive optical pulses with different driving frequencies and amplitudes. 
The simulation starts  from a ground state with $U=0.5$ and the average distance between electrons $\bar d\sim 6.5$.  A pulse  with $\omega_{d1}/\omega_0=0.9$ is switched on that gives  rise to an attractive interaction resulting in a decrease of   $\bar d $ as shown in Fig.~\ref{Fig4S} d) while  $g(j,t)$ peaks around  $j=0$, as consistent with the increase of double occupancy.  The second pulse with $\omega_{d2}/\omega_0=1.1$ results in the increase of $\bar d$  indicating the change of the interaction from attractive to repulsive. Both pulses lead to an increase of the total as well as the kinetic energy, $E$ and $E_\mat{kin}$, respectively, as seen in Fig.~\ref{Fig4S} b) and a slight drop of the EP coupling energy $E_{g2}$. 

\section{The evolution of  $g(j)$ with U}

\begin{figure}[!tbh]
\includegraphics[width=1.0\columnwidth]{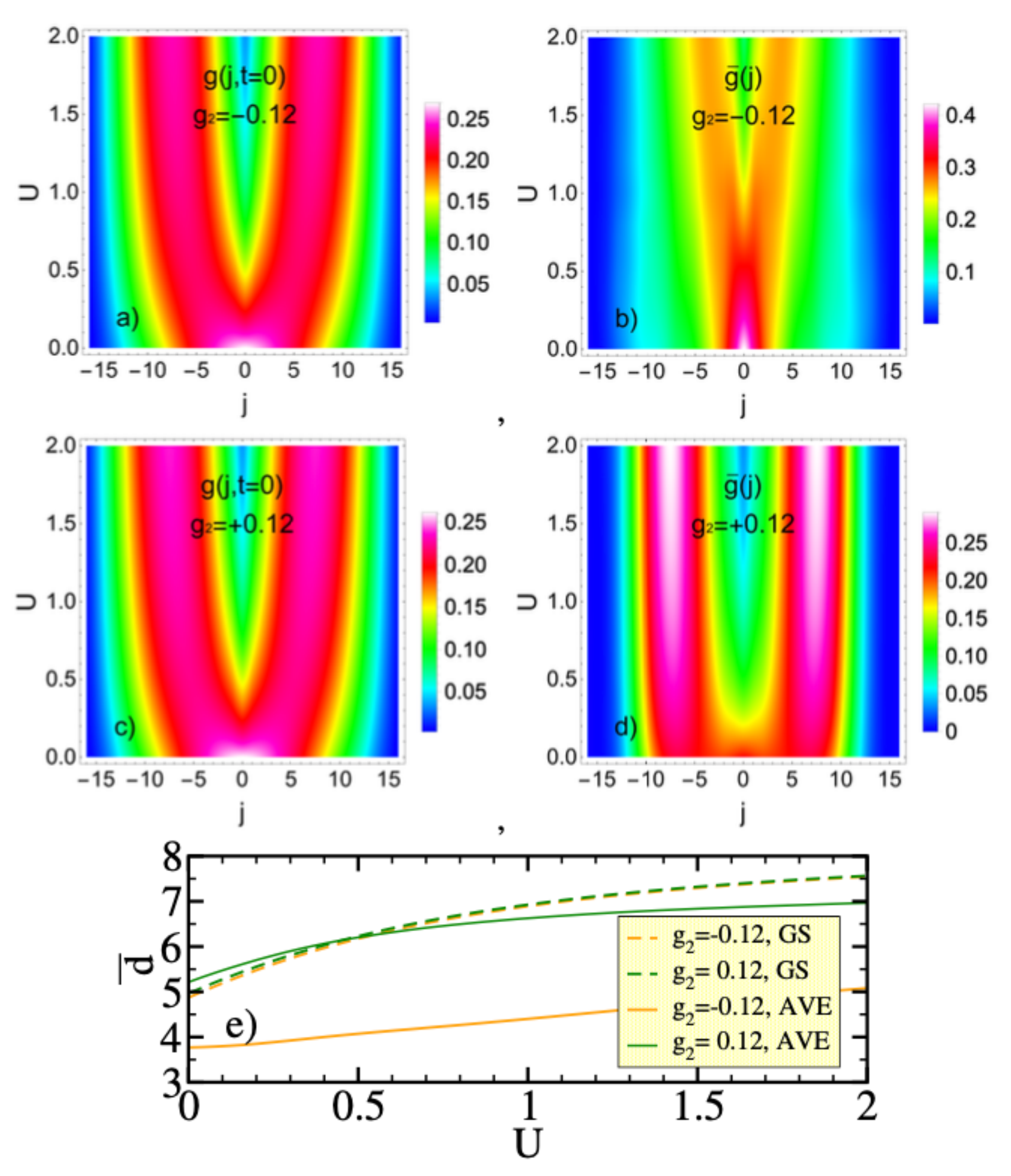}
\caption{ 
 a) and c):  the ground-state $g(j,t=0)$ vs. $U$ for $g_2=-0.12$ and 0.12, respectively; b) and d) time-averaged $\bar g(j)$ for  $g_2=-0.12$ and 0.12, respectively. In the latter case, we have used  driving frequencies $\omega=0.9$ and 1.0, respectively, where maximal attractive interaction is expected based on Figs.~3a) and c) of the mani text; e) corresponding average distance $\bar d$, computed in the ground state (GS) and time-averaged after the pulse (AVE). Note that results for $g_2=\pm0.12$ in the ground state nearly overlap. 
}
\label{Fig9S}
\end{figure}

In Figs.~\ref{Fig9S} a) and c)  we first present the evolution of $g(j,t=0)$ in the ground state with increasing $U$ for $g_2=-0.12$ and 0.12, respectively. Even though the two plots seem identical, there are subtle differences since the Hamiltonian is not invariant to the sign  change of $g_2$.  In Figs.~\ref{Fig9S} b) and d) we present $\bar g(j)$ vs. $U$, time-averaged after the driving pulse for the same values of $g_2$ as in a) and c).  While $g_2>0$ tends to stabilise particles at large distances  after driving, $g_2<0$, in contrast, tends to  weaken repulsive $U$ even at large $U=2$ where $\bar g(j)$ reaches its maximal value around $j\sim 4$, which is in contrast to $j\sim 8$ in the ground state. 

In  Fig.~\ref{Fig9S}e) we present  average distance $\bar d$   in the ground state for two $g_2=\pm 0.12$ as well as after driving. For $g_2=-0.12$ the result clearly shows substantially diminished $\bar d$ in comparison to its value in the ground state. We now focus to results obtained using $g_2=0.12$.  By  comparing Figs.~\ref{Fig9S}c) and  d) we see a clear redistribution of $\bar g(j)$ in comparison with $g(j,t=0)$.  The average distance $\bar d$ in Figs.~9e) for $g_2=0.12$ show a slight increase of $\bar d$ after driving for $U<0.5$ and a slight decrease for $U>0.5$. Still, changes are too small for results to be conclusive. Note also, that for $g_2=0.12$ we have chosen $\omega_d$ that seemed optimal to achieve attractive interaction.

\section{Driven single-site problem (analytic solution)}

Numerical simulations show a resonant behavior of the optically driven system. We analyse the driving process analytically in the atomic limit ($t=0$). The starting point is the Hamiltonian
\begin{multline} \label{H_quadratic_driven_atomic}
	\mathcal{H}_{at} =   U\hat{n}_{\uparrow}\hat{n}_{\downarrow} + \omega_0\Big(b^\dagger b + \frac{1}{2}\Big) + g_2\hat{n}\Big(b^\dagger + b\Big)^2 \\
	+ F(t)\Big(b^\dagger + b\Big).
\end{multline}
First, we apply the squeezing transformation \mbox{$\mathcal{\tilde{H}}_{at}=e^{\hat{S}_1}\mathcal{H}_{at}e^{-\hat{S}_1}$} proposed  by Kennes et al. \cite{millis_2017} which rescales the position and momentum of oscillators depending on a local electronic density. The generator of squeezing transformation can be written as
\begin{equation} \label{squeezing_transf}
	\hat{S}_{1} = \frac{i}{2}\zeta\big(\hat{x}\hat{p} + \hat{p}\hat{x} \big) = -\frac{1}{2}\zeta_j\Big({b^\dagger}^2 - {b}^2\Big),
\end{equation}
with the squeezing parameter $\zeta = -\frac{1}{4}\ln\Big(1 + \frac{4g_2\hat{n}}{\omega_0}\Big)$. \\
Bosonic operators are converted into 
\begin{equation}  \label{mapping_prva_transf}
	\begin{split}
		&b^\dagger\quad\longrightarrow\quad\tilde{b}^\dagger = b^\dagger\cosh\zeta + b\sinh\zeta \\
		&b\quad\longrightarrow\quad\tilde{b} = b\cosh\zeta + b^\dagger\sinh\zeta
	\end{split}
\end{equation}
while the electron number operator remains unaffected. \\
Hamiltonian takes the following form 
\begin{multline} \label{hamiltonka_po_prvi_transf}
	\mathcal{\tilde{H}}_{at} =
	U\hat{n}_{\uparrow}\hat{n}_{\downarrow} + \omega(\hat{n})\Big(b^\dagger b + \frac{1}{2}\Big) \\+ F(t)\Big(1 + \frac{4g_2\hat{n}}{\omega_0}\Big)^{-\frac{1}{4}}\Big(b^\dagger + b\Big),
\end{multline}
where $\omega(\hat{n}) = \omega_0\sqrt{1 + \frac{4g_2\hat{n}}{\omega_0}}$ denotes the electron-density dependent oscillator frequency. \\
Since $\mathcal{\tilde{H}}_{at}$ has a similar form as the single-site Holstein-Hubbard Hamiltonian we proceed with a transformation analogous to the Lang-Firsov transformation. Based on the form of the Lang-Firsov transformation generator we construct 
\begin{equation}
	\hat{S}_2 = \frac{F(t)}{\omega_0}\Big(1 + \frac{4g_2\hat{n}}{\omega_0}\Big)^{-\frac{3}{4}}\Big(b^\dagger - b\Big).
\end{equation} 
This transformation maps bosonic operators 
\begin{align}
	\begin{split}
		&b^\dagger \quad\longrightarrow\quad b^\dagger - \frac{F(t)}{\omega_0}\Big(1 + \frac{4g_2\hat{n}}{\omega_0}\Big)^{-\frac{3}{4}} \\
		&b \quad\longrightarrow\quad b - \frac{F(t)}{\omega_0}\Big(1 + \frac{4g_2\hat{n}}{\omega_0}\Big)^{-\frac{3}{4}}
	\end{split}
\end{align}
and yields transformed Hamiltonian of the form 
\begin{equation} \label{hamiltonian_diagonalised}
	\mathcal{H}_{at}' =   U\hat{n}_{\uparrow}\hat{n}_{\downarrow} + \omega(\hat{n})\Big(b^\dagger b + \frac{1}{2}\Big) - \frac{F(t)^2}{\omega_0}\Big(1 + \frac{4g_2\hat{n}}{\omega_0}\Big)^{-1}.
\end{equation}
The electron number operator is once again unaffected by the transformation. 
\vspace{2mm}

\noindent
The dependence of the driving response should be reflected in the effective electron-electron interaction which we introduce following Kennes et al. \cite{millis_2017} as  
\begin{equation}
	U_{eff} = (E_2 - E_0) - 2(E_1 - E_0) = E_2 - 2E_1  + E_0,
\end{equation} 
Considering $\mathcal{\tilde{H}}_{at}$ (we find it more convenient to obtain time-dependencies using $\mathcal{\tilde{H}}_{at}$ instead of $\mathcal{H}_{at}'$) the effective Coulomb interaction takes the following form 
\begin{equation} \label{Ueff}
	\begin{split}
		U_{eff} = &\ U + \omega(2)\langle b^\dagger b\rangle_2 + \omega_0\langle b^\dagger b\rangle_0 - 2\omega(1)\langle b^\dagger b\rangle_1 \\ &  + \frac{1}{2}\Big[\omega(2) + \omega_0 - 2\omega(1)\Big] + F(t)\Big[\gamma(2)\langle b^\dagger + b\rangle_2 \\& + \langle b^\dagger + b\rangle_0 - 2\gamma(1)\langle b^\dagger + b\rangle_1 \Big],
	\end{split}
\end{equation}
where $\langle b^\dagger b\rangle_n$ and $\langle b^\dagger + b\rangle_n$ denote the expectation value of the $b^\dagger b$ and $b^\dagger + b$ operators in the propagated state with $n\in\{0, 1, 2\}$ being the electron occupation number. We introduced $\gamma(\hat{n}) = \Big(1 + \frac{4g_2\hat{n}}{\omega_0}\Big)^{-\frac{1}{4}}$ to obtain expression in a more compact form. The electron occupation number operators in $\omega(\hat{n})$ and $\gamma(\hat{n})$ can be replaced with scalars in the atomic limit. \\
We believe that there is a strong correlation between the behavior of $U_{eff}$ and the expectation number of phonons excited during the driving. Bearing in mind the mapping of bosonic operators \eqref{mapping_prva_transf} the phonon number operator is written as 
\begin{multline}
	\hat{N}_{ph}(t) = b^\dagger(t)b(t)\Big(\cosh^2\zeta(n) + \sinh^2\zeta(n)\Big) + \sinh^2\zeta(n) \\+ \Big(b^\dagger(t)^2 + b(t)^2\Big)\sinh\zeta(n)\cosh\zeta(n)
\end{multline}
We calculate time evolutions of $U_{eff}$ and $\langle \hat{N}_{ph}\rangle_n$ using the Heisenberg picture. For simplicity we assume the harmonic driving  $F(t) = F_0\sin(\omega_dt)$. The Heisenberg equations of motion with respect to \eqref{hamiltonka_po_prvi_transf} yield 
\begin{equation}
	\begin{split}
		b(t) = be^{-i\omega(n)t} - \frac{iF_0\gamma(n)\omega_d}{\omega(n)^2 - \omega_d^2}\cos(\omega_dt) \\- \frac{F_0\gamma(n)\omega_d}{\omega(n)^2 - \omega_d^2}\sin(\omega_dt), \\
		b^\dagger(t) = b^\dagger e^{i\omega(n)t} + \frac{iF_0\gamma(n)\omega_d}{\omega(n)^2 - \omega_d^2}\cos(\omega_dt) \\- \frac{F_0\gamma(n)\omega_d}{\omega(n)^2 - \omega_d^2}\sin(\omega_dt),
	\end{split}
\end{equation}
The initial state of the system is a coherent state $\ket{\alpha}$ with $\langle \hat{x}(t=0)\rangle = 0$ and $\langle \dot{\hat{x}}(t=0)\rangle = 0$. Hence $\Re(\alpha) = 0$ and $\Im(\alpha) = \frac{F_0\gamma(n)\omega_d}{\omega(n)^2 - \omega_d^2}$, where $\alpha$ denotes the eigenvalue of $b$. It follows 
\begin{equation}
	\begin{split}
		\langle b^\dagger(t)b(t) \rangle_n = \frac{2F_0^2\gamma(n)^2\omega_d^2}{(\omega(n)^2 - \omega_d^2)^2}\Big[1 - \cos(\omega_dt)\cos(\omega(n)t) \\- \sin(\omega_dt)\sin(\omega(n)t)\Big], \\
		\langle b^\dagger(t) + b(t) \rangle_n =  \frac{2F_0\gamma(n)\omega_d}{\omega(n)^2 - \omega_d^2}\Big[\sin(\omega(n)t) - \sin(\omega_dt) \Big], \\
		\langle b^\dagger(t)^2 + b(t)^2 \rangle_n = \frac{2F_0^2\gamma(n)^2\omega_d^2}{(\omega(n)^2 - \omega_d^2)^2}\Big[\sin^2(\omega_dt) - \cos^2(\omega_dt) \\- \cos(2\omega(n)t) + 2\cos(\omega_dt)\cos(\omega(n)t)\\ - 2\sin(\omega_dt)\sin(\omega(n)t)\Big], 
	\end{split}
\end{equation}
which determines the time-evolution of $\langle \hat{N}_{ph}\rangle_n$ and $U_{eff}$. 

\begin{figure}[!tbh]
	\centering
	\includegraphics[width=1.\linewidth]{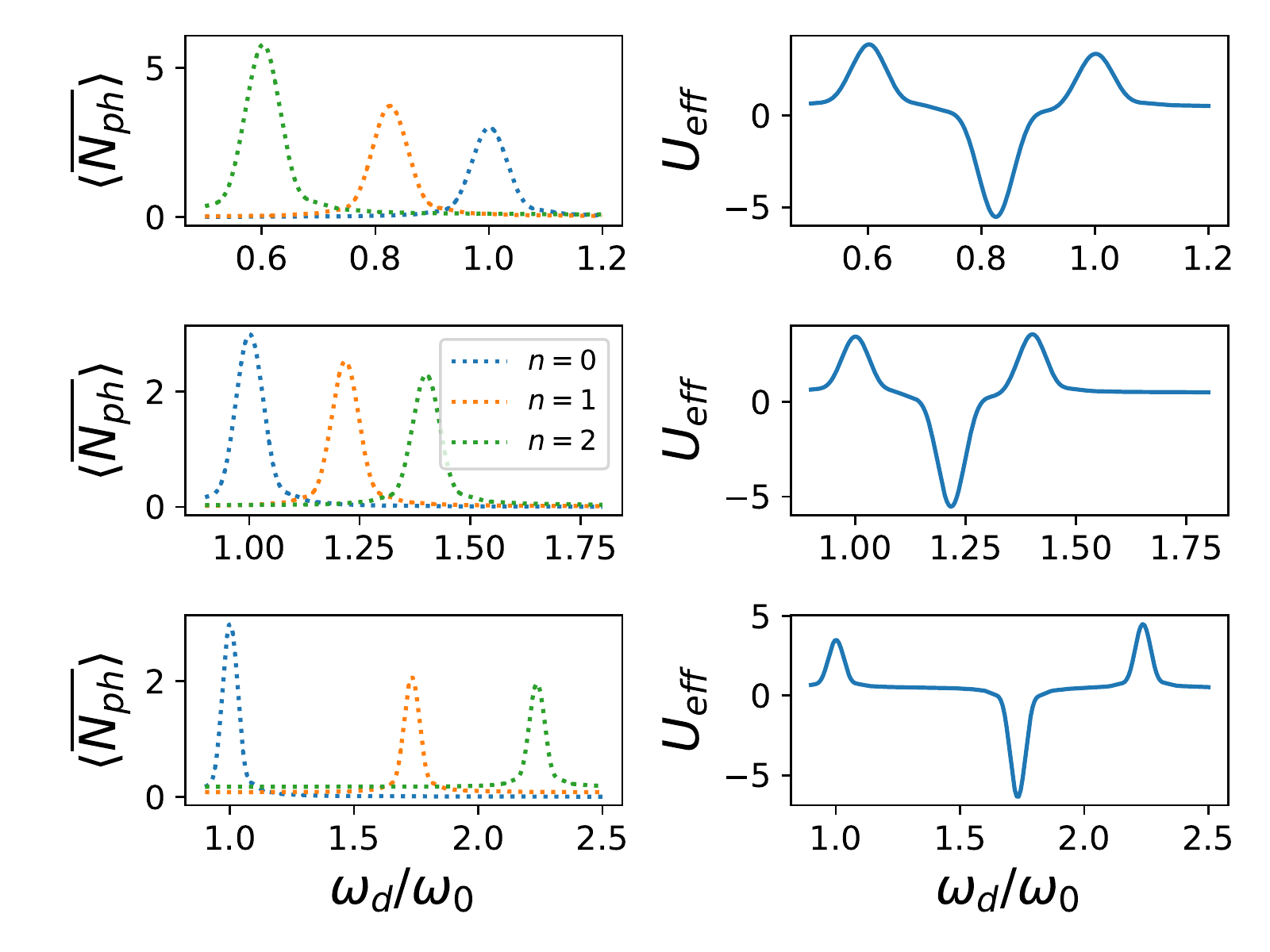}
	\caption{Time-averaged number of phonons and effective potential as a function of the driving frequency. Each row corresponds to a different value of $g_2$. The upper one to $g_2 = -0.08$, the middle one to $g_2 = 0.12$ and the bottom one to $g_2 = 0.5$.
	}
	\label{Fig10S}
\end{figure}

In Fig.~\ref{Fig10S} we plot the time-averaged number of phonons along with the time-averaged potential. We have computed $\langle b^\dagger(t)b(t) \rangle_n$, $\langle b^\dagger(t) + b(t) \rangle_n$ and $\langle b^\dagger(t)^2 + b(t)^2 \rangle_n$ for each $n=0,1,2$ as a function of the driving frequency $\omega_d$. These values determine the $\omega_d$-dependence of $\langle \hat{N}_{ph}\rangle_n$ and $U_{eff}$. 
%The system was driven for 100 time units by time--steps  0.1. The average number of phonons was then obtained by dividing the sum of collected $\langle \hat{N}_{ph}\rangle_n$ with the number of samples. The effective potential as a function of the driving frequency was calculated by putting these averages into Eq. \eqref{Ueff}. 
We observe a resonant behavior at frequencies that differ from bare molecular oscillator frequency depending on the electron occupation number and the value of the electron-phonon coupling parameter. At frequencies close to $\omega_0$ and $\omega(2)$ effective Coulomb interaction increases, however, there is also a strong dip in $U_{eff}$ centered at $\omega(1)$. 

The most important contribution to $U_{eff}$ comes from $\langle b^\dagger b\rangle_n$. This term would persist even if the driving stops and has the same resonant behavior as the number of phonons that get excited during the driving process. 

\section{Physical realizations of quadratic coupling}

In this section, we gather different physical realization where quadratic electron-lattice coupling is dominant. 

The first example includes materials with dipolar active phonons in a centrosymmetric system where the linear term is forbidden by symmetry. This is the experimentally relavant example in  K$_3$C$_{60}$\cite{cavalleri_Nat2016}or $\kappa$-(ET)$_2$Cu[N(CN)$_2$]Br ($\kappa$-Br)\cite{cavalleri_PRX2022}. 

\begin{figure}[!tbh]
	\centering
	\includegraphics[width=1.\linewidth]{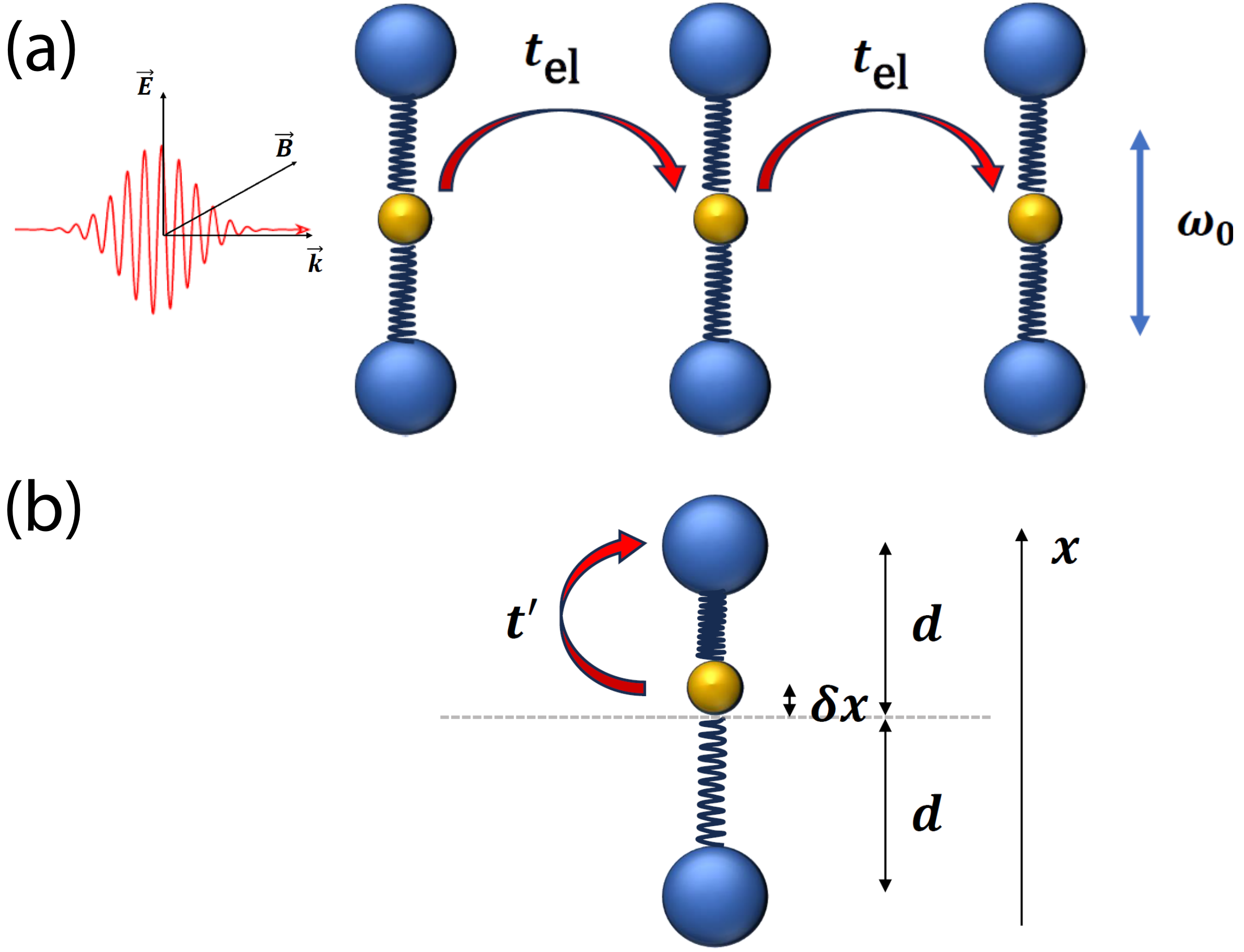}
	\caption{a) Sketch of a setup with a low-dimensional system where light ions are symmetrically intercalated between heavy ions and the electric field whose polarization is applied perpendicular to the direction of the lattice. b)~Sketch of the displacement.}
	\label{Fig11S}
\end{figure}

The second example includes systems formed by a chain of molecular oscillators, each made up of three atoms, with light atoms depicted as yellow spheres and positioned between the heavy ones represented as blue spheres, in Fig.~\ref{Fig11S}(a). As an initial approximation, the motion of the heavy atoms can be neglected, while vibrations of the light atoms are described by independent harmonic oscillators with frequency $\omega_0$ corresponding to optical phonons. As these systems are low-dimensional, we can reduce the heating by using polarization which is perpendicular to the active dimension of the system. While details of the electron-lattice coupling depend on the actual material, we can consider two main contributions: Coulomb interaction between heavy and light nuclei and tunneling terms. 

Our approach closely follows derivation given in Ref.~\cite{berciu2014a} (Ref.~$^{35}$ in the main manuscript). The first contribution stems from an additional Coulomb interaction between the heavy and light ions due to the presence of the carrier, altering the total charge of the light ion. If we set $U(x)$ as the supplementary Coulomb interaction induced by the carrier, the potential increases by $U(d + \delta x) + U(d - \delta x)$, where $d$ denotes the equilibrium distance between the light and heavy ions, see Fig.~\ref{Fig11S}. This function is even with respect to $\delta x$, implying an absence of linear or odd terms in $\delta x$. Consequently, the interaction component of the Hamiltonian can be expressed as $\mathcal{H}_{int} \propto \sum_j n_j \delta x_j^2 \propto \sum_j n_j (a_j^\dagger + a_j)^2$, where $a_j^\dagger$ ($a_j$) represents an operator that creates (annihilates) a phonon on the $j$-th harmonic oscillator, and $n_j$ signifies the electron density operator of the $j$-th site.

The second mechanism, termed hybridization, explains the EP coupling term in the Hamiltonian as a result of virtual hopping processes within the  second-order perturbation theory. The overlap of orbitals of light and heavy ions is represented by the hopping integral $t'(x)$, depending on the distance between the ions. The carrier can decrease its on-site energy by $-t'^2/\Delta$ via virtual hopping to a neighboring heavy ion and back. Here, $\Delta$ represents the energy difference between the carrier residing in the light ion's orbital and an electron occupying the heavy ion's orbital. We assume $\Delta$ to be substantial, resulting in a negligible probability of the carrier occupying the heavy ion. In instances of small $\delta x$, the hopping can be approximated to the lowest order as $t_0(1 + \alpha\delta x)$, where $t_0 = t'(x = d)$, and $\alpha$ represents a material-specific constant. Two contributions emerge from the carrier's ability to hop to either of the two heavy ions, amounting to
\begin{equation}
	\frac{-t_0^2}{\Delta}\big[(1 + \alpha\delta x)^2 + (1 - \alpha\delta x)^2\big] = \frac{-2t_0^2}{\Delta}\big[1 + \alpha^2\delta x^2\big]
\end{equation}
Given that the change in energy manifests as a quadratic function of $\delta x$, it follows that $\mathcal{H}_{int} \propto \sum_j\hat{n}_j(a_j^\dagger + a_j)^2$ and
importantly the prefactor of such contribution is negative relevant for photo-induced pairing introduced in the main text.

\newpage

\bibliography{manuaomm}